\pdfoutput=1

\documentclass[sigconf,nonacm]{acmart}

\setcopyright{none}
\renewcommand\footnotetextcopyrightpermission[1]{}

\usepackage{packages/packages}
\usepackage{packages/macros}

\begin{document}

\title{\ours: Co-Optimizing RAG Serving Performance and Quality}

\author{Haiqiang Zhang}
\affiliation{%
  \institution{ETH Zurich}
  \city{Zurich}
  \country{Switzerland}
}
\email{zhaiqiang@ethz.ch}

\author{Yuanqing Lei}
\affiliation{%
  \institution{Columbia University}
  \city{New York}
  \country{United States}
}
\email{yl5457@columbia.edu}

\author{Wanting Li}
\affiliation{%
  \institution{National University of Singapore}
  \city{Singapore}
  \country{Singapore}
}
\email{wantingli@u.nus.edu}

\author{Tao Zhang}
\affiliation{%
  \institution{ETH Zurich}
  \city{Zurich}
  \country{Switzerland}
}
\email{zhangta@ethz.ch}

\author{Wenqi Jiang}
\affiliation{%
  \institution{National University of Singapore}
  \city{Singapore}
  \country{Singapore}
}
\email{wenqi.jiang@nus.edu.sg}

\renewcommand{\shortauthors}{Zhang et al.}

\begin{abstract}
Retrieval-augmented generation (RAG), which augments LLM generation with information retrieved from databases, has become a widely used approach for knowledge-intensive applications.
Modern RAG systems, however, expose many configuration choices, such as retrieval indexes, model selections, and how models invoke retrievals.
Each configuration yields a different trade-off between answer quality and serving performance, making it challenging to choose the optimal setting for a specific application deployment.
In this paper, we present \ours{}, a framework for efficiently discovering quality--performance Pareto frontiers across diverse RAG applications and serving systems.
Specifically, it consists of \plan{}, an iterative design-space exploration algorithm that selects the next RAG configuration to evaluate; \ir{}, a workload abstraction for diverse RAG algorithms; and \perf{}, a performance model that predicts the optimal deployment and serving performance on the given hardware.
Together, these components allow \ours{} to search the joint algorithm--system configuration space without deploying every candidate and to transfer an existing Pareto frontier to a new serving system.
Given the same number of optimization iterations across diverse datasets, the Pareto frontiers found by \ours{} cover 52.5--153.2\% more of the normalized quality--performance space than those found by state-of-the-art configuration-search methods evaluated over the same RAG design space.

\end{abstract}

\maketitle
\pagestyle{plain}

\vspace{.3cm}
\begingroup\small\noindent\raggedright\textbf{Artifact Availability:}\\
The source code, data, and/or other artifacts have been made available at
\url{https://github.com/haiqiang-zhang/rag-stack}.
\endgroup

\section{Introduction}
\label{sec:intro}

Retrieval-augmented generation (RAG) augments large language models (LLMs) with information retrieved from external data sources~\cite{vuFreshLLMsRefreshingLarge2024,lewisRetrievalaugmentedGenerationKnowledgeintensive2020, gaoRetrievalAugmentedGenerationLarge2024}.
By combining models and databases, RAG addresses several limitations of model-only systems.
First, RAG can incorporate up-to-date information that may not be encoded in a model's parameters~\cite{vuFreshLLMsRefreshingLarge2024}.
Second, by grounding generation in retrieved evidence, RAG can improve factual accuracy and reduce hallucinations~\cite{shusterRetrievalAugmentationReduces2021}.
Third, RAG enables models to use private or domain-specific corpora that were unavailable during pretraining~\cite{zhangRAG4ITOpsSupervisedFineTunable2024, siriwardhanaImprovingDomainAdaptation2023}.
Together, these capabilities have made RAG a widely adopted approach for knowledge-intensive applications~\cite{zhangRAG4ITOpsSupervisedFineTunable2024, zhuRAGEvalScenarioSpecific2025, chenFinTextQADatasetLongform2024}.

\begin{figure}[t]
  \centering
  \includegraphics[width=\columnwidth]{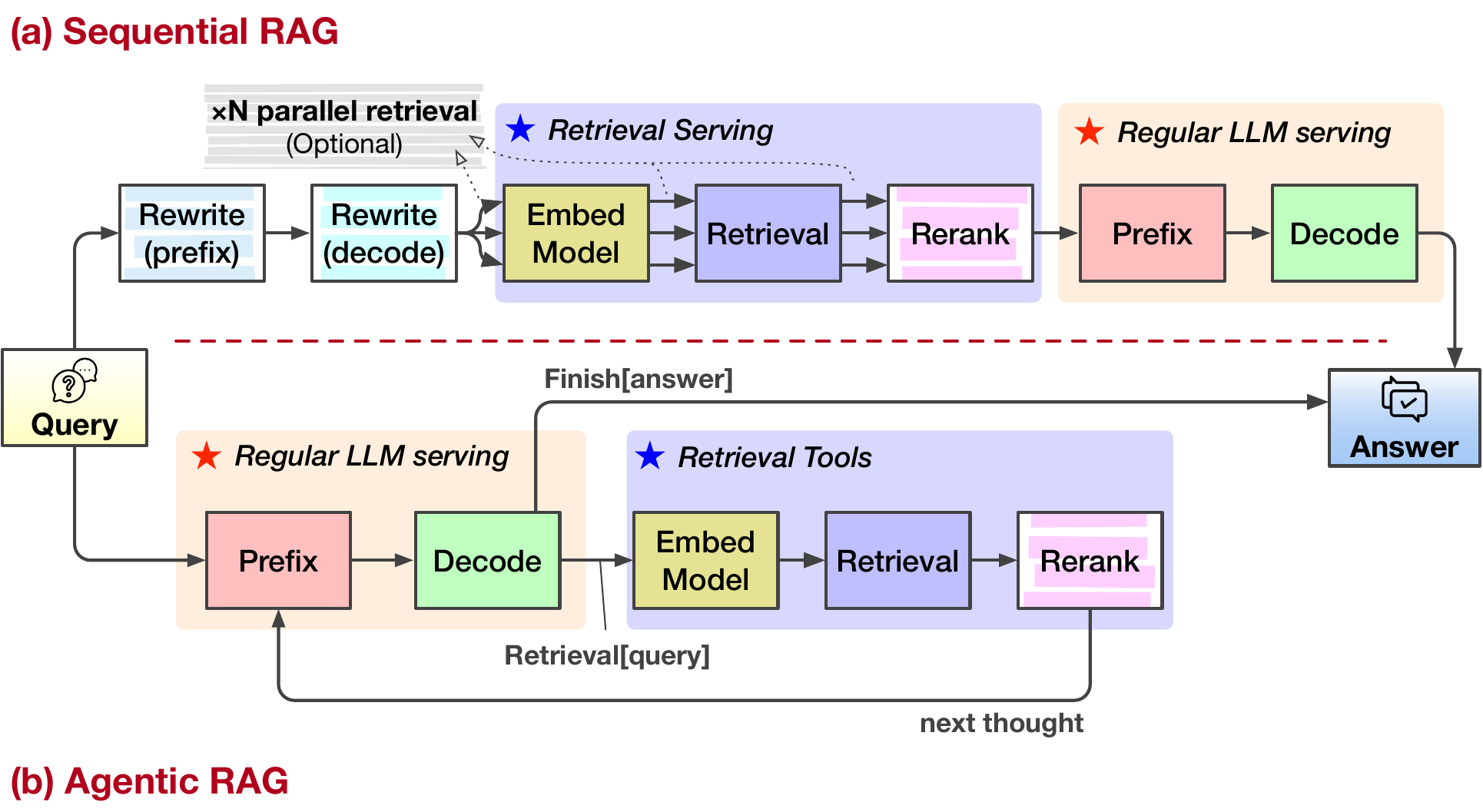}
  \vspace{-2em}
  \caption{Modern RAG system comprises diverse model components and execution workflows.}
  \vspace{-2em}
  \label{fig:pipeline}
\end{figure}

Although the core idea of RAG is simple, modern RAG systems expose a large design space spanning both (1) pipeline components and (2) execution workflows (Figure~\ref{fig:pipeline}), inducing a broad spectrum of trade-offs between answer quality and serving performance.
At the component level, a RAG pipeline contains a retrieval system and a generative LLM and may additionally include stages such as query rewriting and result reranking.
Each RAG component can introduce a unique choice of model, prompt, and stage-specific parameters.
At the workflow level, these components can be orchestrated either as a \emph{(a) sequential pipeline}, in which every query follows a fixed execution order, or as an \emph{(b) agentic pipeline}, in which an LLM-based controller dynamically decides when to retrieve and whether additional retrieval rounds are needed~\cite{liAgenticRAGDeep2025}.
Naively enabling every pipeline component, selecting the most capable models, or maximizing the number of retrieval rounds may improve answer quality, but doing so also increases serving cost and reduces system performance (worse latency and throughput).
%
%
In this paper, we ask: \textit{How can we efficiently explore this quality--performance trade-off space and identify the Pareto frontier across diverse RAG applications and serving systems?}

However, identifying the Pareto frontier for a given RAG application and serving system requires solving three key problems.
\textbf{(P1) Existing RAG configuration exploration algorithms either neglect cross-stage interactions or overlook per-stage signals.}
One approach to explore RAG configurations is to proceed stage by stage: it identifies the best configuration for one stage, holds that configuration fixed, and then optimizes the next stage~\cite{MarkerIncKoreaAutoRAG2026}.
However, the best end-to-end configuration depends on interactions among stages and therefore cannot generally be obtained by optimizing each stage independently.
For example, increasing the retrieval top-$k$ can improve quality when a strong reranker filters irrelevant passages or a long-context LLM effectively uses the additional evidence, but can degrade quality when noisy passages reach a weaker generator without reranking.
By contrast, global RAG optimization captures these interactions by jointly tuning parameters across all stages using end-to-end accuracy as the optimization signal~\cite{barkerFasterCheaperBetter2025a}.
However, this end-to-end signal does not reveal which stage is responsible for a gain or loss, potentially leading to poorly directed exploration and wasted evaluations.
\textbf{(P2) Existing multi-objective RAG optimizers overlook the system design space.}
Existing quality--performance optimizers search over algorithmic configurations while holding the serving deployment fixed~\cite{barkerFasterCheaperBetter2025a, rayMETISFastQualityAware2025}, even though different algorithms may require different batching policies, parallelization strategies, and stage placements to achieve their best serving performance~\cite{jiangRAGOSystematicPerformance2025}.
%
\textbf{(P3) Deployment-based serving-performance measurements are costly and do not transfer across systems.}
Measuring serving performance requires deploying and benchmarking each candidate configuration, and the resulting measurements are specific to the target system~\cite{liRAGPerfEndtoEndBenchmarking2026}.
When the application is moved to a system with different hardware, configurations on the original Pareto frontier may no longer be Pareto-optimal, requiring the search to be repeated to reconstruct the frontier~\cite{luNeuralArchitectureTransfer2021, jamshidiLearningSampleExploiting2018}.

\begin{figure}[t]
  \centering
  \includegraphics[width=\columnwidth]{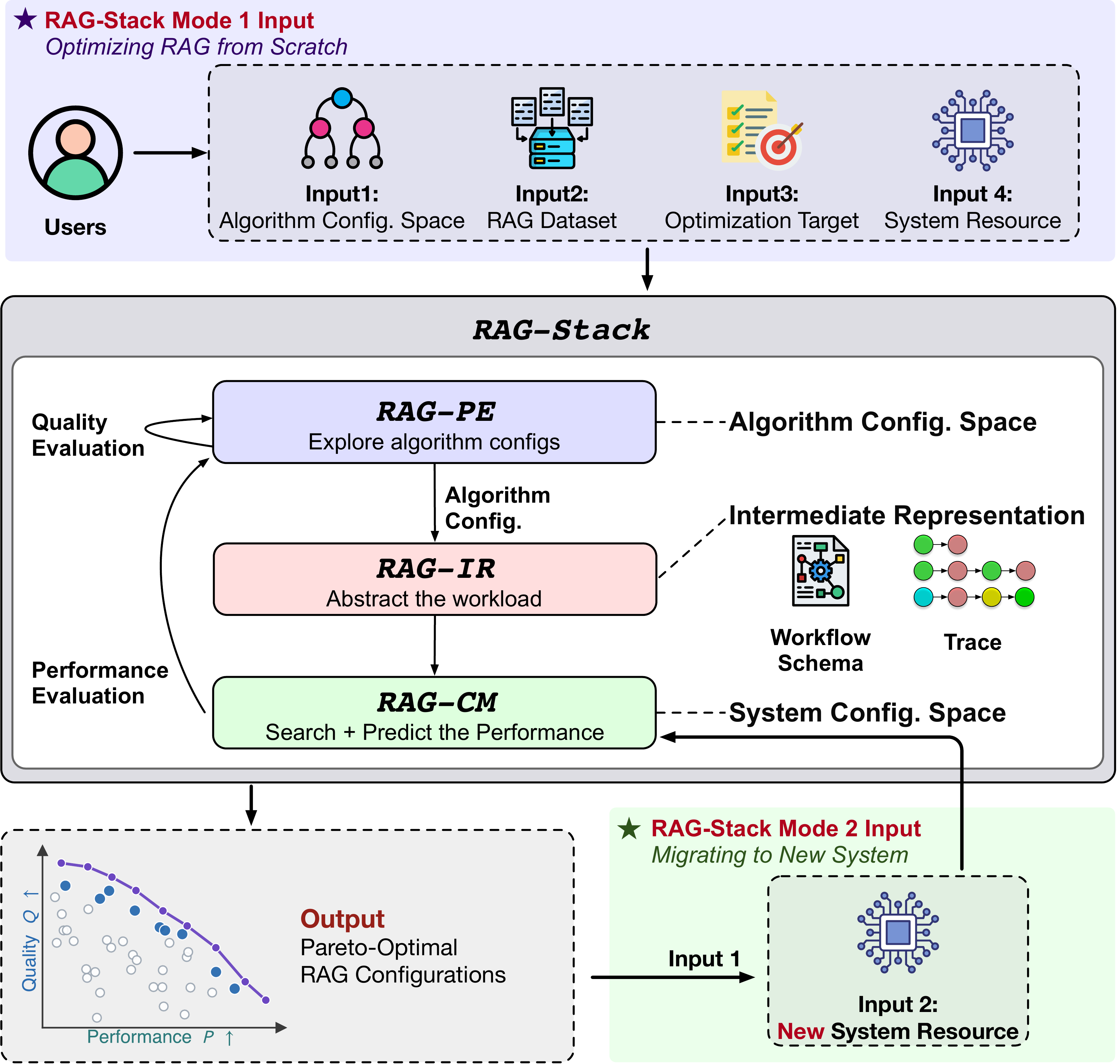}
  \vspace{-2em}
  \caption{Overview of \ours{} and its two operating modes.}
  \vspace{-2.5em}
  \label{fig:overview}
\end{figure}

To address these problems, we present \ours, an efficient framework for discovering the quality--performance Pareto frontier across arbitrary RAG applications and serving systems (Figure~\ref{fig:overview}).
\ours takes four inputs from the user:
(1) an algorithm design space,
(2) a new RAG application (i.e., an evaluation dataset),
(3) an optimization target comprising quality and performance requirements, and
(4) the available system resources.
\ours can either discover a Pareto frontier from scratch or transfer an existing frontier to a new system by reusing archived quality measurements.

For a new RAG application, \ours discovers the Pareto frontier with an iterative optimization loop.
First, \plan proposes an algorithm configuration for quality evaluation on the application dataset.
Next, \ir abstracts the executed workflow, and \perf predicts its best achievable serving performance under the available resources without requiring deployment on the target system.
Finally, \plan uses the measured quality and predicted performance to select the next configuration.
This loop continues until a user-specified stopping criterion or search budget is reached.
We now introduce the main components in \ours.

First, \plan (\underline{P}lan \underline{E}xploration) is the core multi-objective Bayesian optimizer in \ours.
It takes as input the algorithm design space and feedback from previous trials, including quality measurements on the dataset and performance estimates produced by \perf.
Its output is the next algorithm configuration to evaluate, which is sent through the RAG execution path and subsequently represented by \ir.
\plan addresses \textbf{P1} by jointly optimizing end-to-end quality and serving performance while using intermediate stage-level quality signals to guide exploration.

Second, \ir (\underline{I}ntermediate \underline{R}epresentation) bridges \plan{} and \perf{} by translating each RAG algorithm configuration into a workload representation comprising a workflow schema and an execution trace of stage invocations and input/output sizes. 
The output of \ir{} is a \emph{system-agnostic} workload representation that decouples each stage’s logical work from its physical deployment, enabling \perf{} to explore deployment choices and re-estimate the performance of the same workload on new hardware.
%
%
Together with \perf, this representation addresses \textbf{P2} and \textbf{P3}.

Third, \perf (\underline{C}ost \underline{M}odel) is an ML--analytical fusion performance model.
It takes as input the workload representation produced by \ir, the system design space, and the user's available hardware resources.
It then searches the system design space internally and returns the best predicted deployment and serving performance for the current RAG configuration to RAG-PE, thus effectively addressing \textbf{P2}.
Moreover, RAG-CM can re-evaluate the archived algorithm configurations under a new hardware configuration without repeating their quality evaluations, enabling efficient frontier transfer and further addressing \textbf{P3}.

We evaluate \ours{} on RAGEval~\cite{zhuRAGEvalScenarioSpecific2025} and MS MARCO~\cite{bajajMSMARCOHuman2018} using two system configurations: one equipped with four NVIDIA H100 GPUs and the other with eight NVIDIA A100 GPUs.
With the same number of exploration iterations, the Pareto frontiers found by \ours{}, averaged across seeds, cover 52.5\% and 153.2\% more of the normalized quality--performance space on RAGEval and MS MARCO, respectively, than those found by state-of-the-art configuration-search methods evaluated over the same RAG design space.
When transferring a Pareto frontier to a new serving system, \ours{} reuses quality measurements from previous evaluations and performs only a small number of additional optimization iterations to discover the new frontier. The resulting adapted frontier covers 182.2\% more of the normalized quality--performance space than the frontier obtained by re-optimizing from scratch on the new system.

In summary, the paper makes the following \textbf{contributions}:
\begin{itemize}[
    leftmargin=1.2em,
    labelsep=0.5em,
    topsep=3pt,
    itemsep=2pt
]
  \item We present \ours, an end-to-end framework that efficiently discovers the Pareto frontier between RAG answer quality and serving performance for arbitrary applications and systems.
  \item We design \plan, a multi-objective, sub-metric-aware Bayesian optimizer that uses both end-to-end and intermediate signals to efficiently navigate the RAG configuration space.
  \item We introduce \perf, a hybrid ML-analytical performance model that predicts serving performance and searches optimal deployment configurations on the given hardware.
\end{itemize}

\section{Background and Motivation}
\label{sec:background}

\subsection{Vector Search for Retrieval}
\label{sec:bg:vector-backends}

RAG retrievers use \emph{vector search} to match queries with passages based on semantic similarity rather than exact lexical overlap. Before serving, corpus passages are embedded and indexed. At query time, the retriever embeds the query and returns the nearest indexed passages. Exact search scans the entire corpus and guarantees the true nearest neighbors under the chosen metric, but scales poorly. Production systems therefore rely on \emph{approximate nearest-neighbor} (ANN) indexes, which prune most candidates to reduce latency and increase throughput, but may lower recall and consequently degrade answer quality when relevant evidence is missed.

\textbf{Two index families.} ANN indexes are either clustering-based or graph-based. The \emph{IVF (inverted-file) family} is clustering-based: it groups the vectors into many lists, and for each query scans only the few lists closest to it. Within this family, IVF-Flat keeps the full vectors and favors recall; IVF-PQ compresses each vector into a short code to save memory and bandwidth; and IVF-PQ FastScan speeds up the distance computation with SIMD-friendly kernels~\cite{jegouProductQuantizationNearest2011,douzeFaissLibrary2025}. \emph{HNSW} is graph-based: it links the vectors into a navigable graph and answers a query by walking the graph greedily toward the nearest ones. HNSW often reaches high recall at low latency, but it needs extra memory to store the graph~\cite{malkovEfficientRobustApproximate2018a}.

\textbf{Knobs trade recall for speed.} Larger \texttt{nprobe} in IVF or \texttt{efSearch} in HNSW examines more candidates, improving recall at the cost of latency and throughput; IVF-PQ reduces memory use but may lower recall. Because the best choice depends on the workload and deployment, FAISS is well suited to studying these trade-offs: it directly exposes both index families and their CPU/GPU knobs~\cite{douzeFaissLibrary2025}, unlike vector databases such as Milvus, whose serving layer hides many low-level choices~\cite{wangMilvusPurposeBuiltVector2021a}.

\subsection{RAG Pipelines and Serving}
\label{sec:bg:rag-pipeline}

A RAG pipeline is built from a few stages. The two core stages are retrieval (\S\ref{sec:bg:vector-backends}) and generation: the retriever runs a vector search over the corpus, and the retrieved passages---optionally after query rewriting or reranking---are placed in a prompt for an LLM that prefills and decodes. The same stages can be wired into different \emph{workflows}. The simplest pipeline retrieves once and generates once. Iterative and active-retrieval workflows interleave several rounds of retrieval and generation to improve the answer~\cite{asaiSelfRAGLearningRetrieve2023}. An agentic pipeline goes further: an LLM controller decides at run time which stages to run, when to retrieve, and whether to iterate~\cite{singhAgenticRetrievalAugmentedGeneration2026,liAgenticRAGDeep2025, bestaAffordableAIAssistants2025a}. Choosing the components and the workflow is already a quality-tuning problem, and frameworks such as FlashRAG let users assemble, swap, and evaluate these pipelines to raise answer quality~\cite{jinFlashRAGModularToolkit2025}.

The choices that raise quality also raise the work per request. A larger top-$k$, an added reranker, iterative retrieval, or a bigger generator each helps the answer, but each also adds retrieval work, model calls, prompt tokens, or decoding time. Serving then adds a second layer of choices that leave the answer unchanged but set how fast it is produced: how the stages are placed on the hardware, how much parallelism they use, how requests are batched, and how the index and runtime knobs are configured.

A large body of systems work optimizes this serving layer for a fixed RAG algorithm---tensor~\cite{shoeybiMegatronLMTrainingMultiBillion2020}, pipeline~\cite{huangGPipeEfficientTraining2019,harlapPipeDreamFastEfficient2018}, and data~\cite{liPyTorchDistributedExperiences2020} parallelism, continuous batching~\cite{yuOrcaDistributedServing2022}, and paged KV-cache management~\cite{kwonEfficientMemoryManagement2023}---together with RAG-specific techniques such as adaptive pipeline parallelism~\cite{jiangPipeRAGFastRetrievalAugmented2025} and disaggregated accelerators~\cite{jiangChameleonHeterogeneousDisaggregated2025}.

\subsection{Motivation: Performance and Quality Co-optimization for RAG}
\label{sec:bg:optimization}

\begin{figure}[t]
  \centering
  \includegraphics[width=\columnwidth]{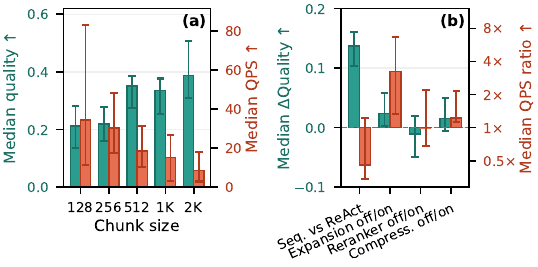}
  \vspace{-2em}
  \caption{Median quality--throughput tension across algorithm choices. (a)~By chunk size. (b)~Single-choice contrasts (quality: first$-$second; QPS: first$/$second); error bars: interquartile range.}
  \vspace{-2em}
  \label{fig:bg-config-selection}
\end{figure}

The discussion above of RAG pipelines and serving shows that co-optimizing answer quality and serving performance better reflects the practical needs of RAG serving. This view also accounts for RAG serving-system design, because placement, batching, and parallelism determine how efficiently each pipeline runs. Manual co-optimization is difficult because measurements reveal trade-offs that intuition misses (Figure~\ref{fig:bg-config-selection}): agentic ReAct~\cite{yaoReActSynergizingReasoning2023} can be faster than a sequential pipeline at a modest quality cost, query expansion sacrifices throughput for a quality gain that appears only in some configurations, and even a larger generator does not always improve quality~\cite{zhangUsingLargeLanguage2026}. A configuration that avoids such unrewarded cost can deliver the same quality at higher throughput and thus \emph{dominate} alternatives; only non-dominated configurations are worth serving.

\phantomsection\label{sec:bg:pareto-search}\textbf{Finding the frontier is a search problem.} Let $\mathcal{X}$ denote a candidate space and $\mathbf{y}(\mathbf{x})\in\mathbb{R}^{m}$ the objective vector of $\mathbf{x}\in\mathcal{X}$, with all objectives oriented so that larger values are better. An objective point $\mathbf{y}$ \emph{dominates} $\mathbf{y}'$ if $\mathbf{y}\ge\mathbf{y}'$ componentwise with at least one strict inequality~\cite{debFastElitistMultiobjective2002}. A candidate is \emph{non-dominated} if no other candidate's objective point dominates its own; the objective points of all non-dominated candidates form the \emph{Pareto frontier} $\mathcal{F}$. The \emph{hypervolume} $\mathrm{HV}(\mathcal{F};\mathbf{r})$ is the volume of the objective-space region that is dominated by $\mathcal{F}$ and dominates a fixed \emph{reference point} $\mathbf{r}\in\mathbb{R}^{m}$ that lower-bounds $\mathcal{F}$; a larger hypervolume indicates broader coverage of desirable trade-offs~\cite{daultonParallelBayesianOptimization2021}. Throughout, objectives are min--max normalized per experiment and dataset, with $\mathbf{r}=\mathbf{0}$. The frontier thus provides a set of operating points from which users can choose. A general configuration-search method finds it iteratively: it proposes a candidate, evaluates its objective vector, and uses the observations collected so far to select the next candidate. Because this search requires only $\mathcal{X}$ and the evaluated objective values, it can in principle use any multi-objective optimization method. The next section shows why such methods, applied directly to RAG, fall short.

\subsection{Limitations of Existing Approaches}
\label{sec:bg:limitations}

Existing RAG-specific optimizers cover only parts of the quality--performance problem. A RAG configuration spans two design spaces: the \emph{algorithm design space}, whose choices affect answer quality---chunking, top-$k$, reranking, and the generator---and the \emph{system design space}, whose choices affect how fast that answer is served---placement, batching, parallelism, and hardware. Quality-side tools search the algorithm design space against answer quality alone~\cite{MarkerIncKoreaAutoRAG2026,jinFlashRAGModularToolkit2025,fuAutoRAGHPAutomaticOnline2024}. System-side tools such as RAGO search the system design space for serving performance, but only for a fixed RAG algorithm~\cite{jiangRAGOSystematicPerformance2025}. Online methods such as METIS adapt a small set of serving choices per query, but they also operate within a fixed deployment~\cite{rayMETISFastQualityAware2025}.

A natural alternative is to treat RAG tuning as a generic multi-objective black-box optimization problem. Recent work takes this route by applying standard multi-objective Bayesian optimization, including LogNEHVI, to search RAG hyperparameters for quality--cost trade-offs~\cite{barkerFasterCheaperBetter2025a}. More broadly, possible black-box optimizers include plain sampling, Bayesian optimization~\cite{lindauerSMAC3VersatileBayesian2022,watanabeTreeStructuredParzenEstimator2026,olsonAxPlatformAdaptive2025,balandatBoTorchFrameworkEfficient2020,levesqueBayesianOptimizationConditional2017}, evolutionary search~\cite{hansenCMAEvolutionStrategy2023,beyerEvolutionStrategiesComprehensive2002}, and multi-fidelity optimization with cheaper proxy runs~\cite{kandasamyMultifidelityBayesianOptimisation2017}. These optimizers can also be steered by LLMs~\cite{laoGPTunerLLMBasedDatabase2025,novikovAlphaEvolveCodingAgent2025,AlgorithmicsuperintelligenceOpenevolve2026} or extended to the multi-objective setting through scalarization~\cite{knowlesParEGOHybridAlgorithm2006,pariaFlexibleFrameworkMultiObjective2019}, dominance~\cite{debFastElitistMultiobjective2002,zhangMOEAMultiobjectiveEvolutionary2007}, or hypervolume criteria~\cite{daultonParallelBayesianOptimization2021,amentUnexpectedImprovementsExpected2025}. These general tools are our starting point. Applied directly to RAG, however, they and the RAG-specific optimizers above run into three problems.

\begin{figure}[t]
  \centering
  \includegraphics[width=\columnwidth]{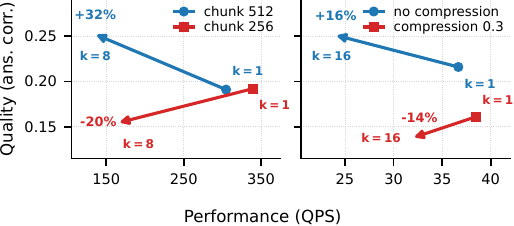}
  \vspace{-2em}
  \caption{Cross-stage interactions in the algorithm design space. Each arrow traces one configuration as retrieval top-$k$ rises; annotated percentages are the resulting quality change.}
  \vspace{-1em}
  \label{fig:bg-interactions}
\end{figure}

\phantomsection\label{sec:bg:lim:p1}\textbf{\underline{(P1)} Existing RAG optimizers either break cross-stage interactions or lose stage-level signals.} Figure~\ref{fig:bg-interactions} shows that increasing retrieval top-$k$ always reduces throughput, but its quality effect flips with chunk size and compression, demonstrating that stage choices cannot be optimized independently. Stage-wise optimizers such as AutoRAG retain stage-level feedback but miss configurations that work only through cross-stage combinations~\cite{MarkerIncKoreaAutoRAG2026}; global optimization accounts for these interactions by tuning complete pipelines, but end-to-end feedback alone cannot reveal which stage drives a gain or loss, leading to poorly directed exploration and wasted evaluations.

\phantomsection\label{sec:bg:lim:p2}\textbf{\underline{(P2)} Existing multi-objective RAG optimizers overlook the system design space.} Existing quality--performance RAG optimizers search algorithm choices but measure performance on one fixed deployment~\cite{barkerFasterCheaperBetter2025a}, while online methods adapt only a few serving parameters within that deployment~\cite{rayMETISFastQualityAware2025}; neither finds an algorithm configuration's best achievable serving performance. Simply folding the full system design space into the multi-objective search is wasteful because system parameters do not affect answer quality, yet every variant would still incur a full quality evaluation.

\phantomsection\label{sec:bg:lim:p3}\textbf{\underline{(P3)} Deployment-based serving-performance measurements are costly and non-transferable across systems.} RAG serving performance is typically obtained by deploying and timing each configuration end to end, making dense deployment search prohibitively expensive~\cite{liRAGPerfEndtoEndBenchmarking2026,liangAthenaPlugandPlayAdvisor2025}. Such measurements are system-specific: a configuration on the original Pareto frontier may become dominated on a new system, so migration requires remeasurement and frontier reconstruction; candidate hardware also cannot be evaluated before acquisition. Although prediction could avoid these costs, existing RAG performance models cover only narrow portions of the stack~\cite{jiangRAGOSystematicPerformance2025,kimVectorLiteRAGLatencyAwareFineGrained2026}, leaving no portable full-stack estimator.

\section{\ours{}: System Overview}
\label{sec:rag-stack}

We present \ours{} (Fig.~\ref{fig:overview}), an efficient framework that finds a RAG system's quality ($Q$)--performance ($P$) Pareto frontier across the full algorithm and system design space, without deploying each candidate to measure its performance. This addresses the limitations in \S\ref{sec:bg:limitations}, which leave RAG without a practical way to co-optimize answer quality and serving performance across its full design space.

\textbf{Inputs.} In from-scratch mode, a user drives \ours{} with a single declarative specification of four things: \textbf{(i)} the \emph{algorithm design space} to search, a hierarchical, conditional space of components and parameters detailed in \S\ref{sec:rag-pe:search-space} (Fig.~\ref{fig:design-space-tree}); \textbf{(ii)} an evaluation \emph{dataset} of queries with reference answers; \textbf{(iii)} an \emph{optimization target}---the quality and performance metrics to trade off, plus serving SLOs; and \textbf{(iv)} the available \emph{system resources}---GPUs, CPU, and interconnect---bounding the system design space \perf{} searches (Table~\ref{tab:system-design-space}). In system-transfer mode, the input is instead the previous Pareto result and a new system-resource specification, as described under \textbf{Two operating modes} below.

\textbf{Output.} \ours{} returns the Pareto-optimal configurations it found---each a complete algorithm-and-system deployment---so the user picks the operating point that fits their workload, hardware, and SLO.

\textbf{Overview.} \ours{} (Fig.~\ref{fig:overview}) rests on one split: a RAG configuration's parameters divide into an \emph{algorithm design space}, whose choices change the produced answer and so move both quality and performance, and a \emph{system design space}, whose choices change only how that fixed computation is served. \ours{} searches the two in an iterative loop over three components. Each round, \plan{} (\underline{P}lan \underline{E}xploration, \S\ref{sec:rag-pe}) proposes one algorithm configuration $\mathbf{x}$; the quality evaluator runs that pipeline on the dataset and returns its quality $Q(\mathbf{x})$; \ir{} (\underline{I}ntermediate \underline{R}epresentation, \S\ref{sec:rag-ir}) abstracts the executed run into a workload representation; \perf{} (\underline{C}ost \underline{M}odel, \S\ref{sec:rag-cm}) searches the system design space for that representation and predicts $\mathbf{x}$'s best serving performance $P(\mathbf{x})$; and \plan{} uses the measured $Q(\mathbf{x})$ and predicted $P(\mathbf{x})$ to choose the next configuration. The loop stops when the optimization target is met or the budget is spent. Quality comes from this one evaluation run, but performance is always predicted by \perf{}---no candidate is ever deployed to measure it.

\textbf{Two operating modes.} The loop above optimizes a RAG system \emph{from scratch}, discovering its frontier on the current hardware. \ours{} also supports a \emph{system-transfer} mode: to retarget an already-optimized system to new hardware, the user provides only the previous Pareto result and a new system-resource specification. \ours{} then reuses the quality results from the first run, first invoking \perf{} to re-score the previous deployments under the new resources and subsequently running a small number of \plan{} polishing iterations to refine the transferred frontier.

\textbf{Benefits.} \ours{}'s design yields three benefits, each removing one limitation of prior optimizers. \textbf{B1 (solving \hyperref[sec:bg:lim:p1]{P1}): stage-aware multi-objective optimization without breaking cross-stage interactions.} \plan{} still optimizes the end-to-end $(Q, P)$ Pareto frontier, so each trial is a complete pipeline configuration and cross-stage interactions are preserved. At the same time, it reads stage-level sub-metrics to guide exploration: for example, low context recall with high faithfulness points to retrieval as the bottleneck rather than generation. These sub-metrics guide where \plan{} searches next without becoming separate objectives, giving one optimizer both global frontier optimization and stage-level diagnosis (\S\ref{sec:rag-pe:optimizer}). \textbf{B2 (solving \hyperref[sec:bg:lim:p2]{P2}): near wallclock-free exhaustive system search.} Given the algorithm/system split above, \plan{} spends expensive quality evaluations only on algorithm configurations, while \perf{} exhaustively searches the system design space for each one inside the cost model. Because this search uses predicted performance rather than real deployments, it is nearly wall-clock-free, letting \ours{} report the best deployment-side operating point instead of inheriting the performance of one fixed deployment (\S\ref{sec:rag-pe:search-space}, \S\ref{sec:rag-cm}). \textbf{B3 (solving \hyperref[sec:bg:lim:p3]{P3}): cheap system transfer and planning.} Because \perf{} predicts performance from a hardware description instead of a real run, an optimized system is retargeted to new hardware by re-scoring the quality archive, and candidate machines are compared before purchase---neither step builds a real deployment.

We now detail the three components in turn: \plan{} (\S\ref{sec:rag-pe}), \ir{} (\S\ref{sec:rag-ir}), and \perf{} (\S\ref{sec:rag-cm}).

\section{\plan{}: Plan Exploration}
\label{sec:rag-pe}


\plan{} partitions the design space (\S\ref{sec:rag-pe:search-space}), runs the RAG-specific optimizer (\S\ref{sec:rag-pe:optimizer}), and coordinates quality evaluation with performance modeling (\S\ref{sec:rag-ir}, \S\ref{sec:rag-cm}).

\subsection{Search Space}\label{sec:rag-pe:search-space}

The space of RAG configurations is large and hierarchical. \plan{} splits it into an algorithm part and a system part, searches the algorithm part directly, and delegates the system part to \perf{}. We define this split and then organize the part \plan{} searches as a hierarchy.

\subsubsection{Two Design Spaces: Algorithm and System Design Spaces.}
\begin{figure*}[t]
  \centering
  \includegraphics[width=0.95\textwidth]{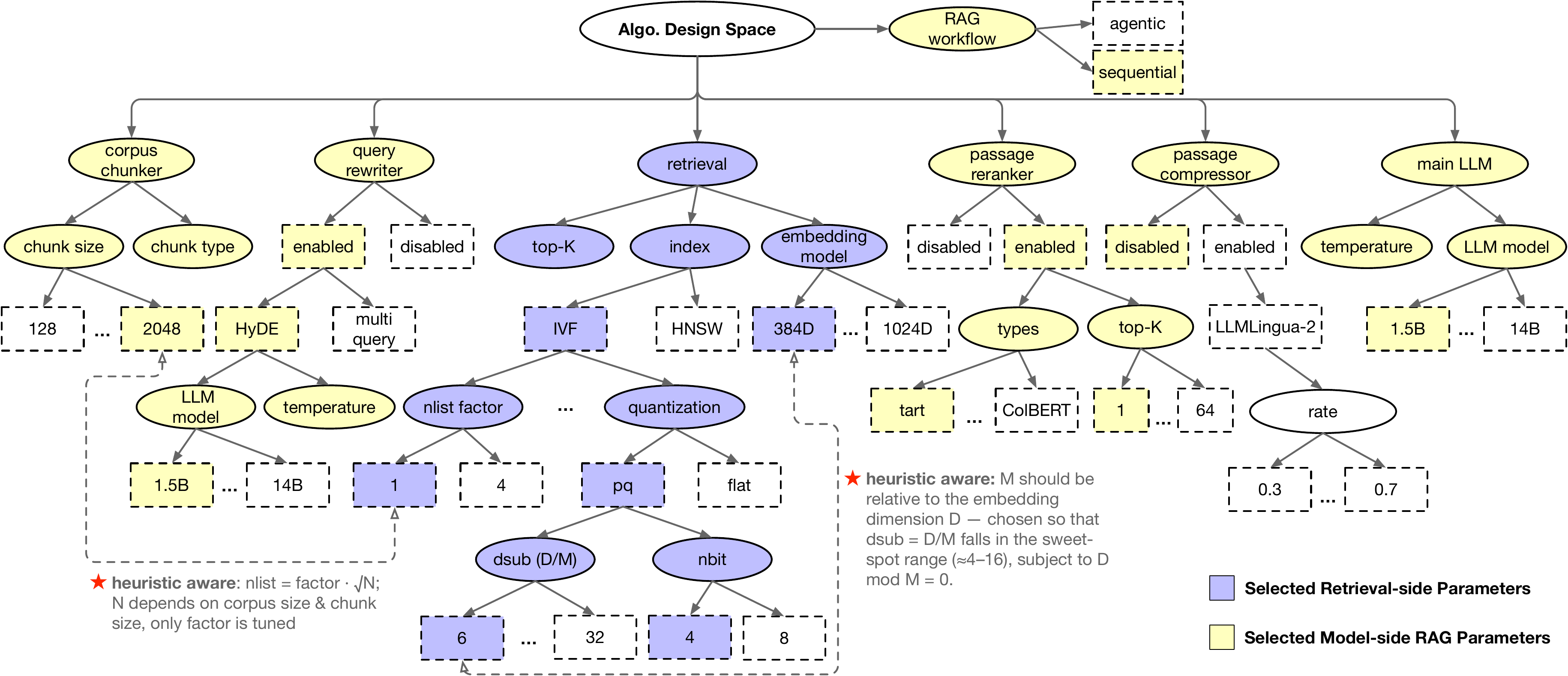}
  \vspace{-1em}
  \caption{Algorithm design space~\cite{panLLMLingua2DataDistillation2024, gaoPreciseZeroShotDense2022, malkovEfficientRobustApproximate2018a,jegouProductQuantizationNearest2011, khattabColBERTEfficientEffective2020a} searched by \plan{}. The highlighted \emph{selected} nodes are one example, showing how a single evaluation's algorithm configuration is assembled: each trial fixes one value at every active branch of the hierarchy (retrieval-side in purple, model-side in yellow), and the starred callouts mark heuristic-aware parameters coupled across branches.}
  \vspace{-1em}
  \label{fig:design-space-tree}
\end{figure*}

Let $\mathcal{X}_{\mathrm{full}} = \prod_{j=1}^{n} \Theta_j$ be the full configuration space of the RAG stack, with each $\Theta_j$ denoting the domain of one parameter (e.g., \texttt{nprobe}, \texttt{thread\_count}, \texttt{top\_k}), and let $Q(\mathbf{x})$ and $P(\mathbf{x})$ be the answer-quality and serving-performance objectives under the user's chosen metrics (e.g., latency, throughput, or SLO satisfaction). We partition the parameters by what each one changes: the \emph{logical computation} that produces the answer, or only the \emph{physical execution} of that fixed computation. An \emph{algorithm parameter} changes the logical computation (Fig.~\ref{fig:design-space-tree}); conversely, \emph{a parameter belongs to the system design space whenever it leaves the logical computation unchanged}. System parameters change only how that fixed computation is executed and served (Table~\ref{tab:system-design-space})---placement, batching, parallelism, thread counts, and hardware. Batching, for example, merely groups identical per-request computations and so leaves the answer unchanged; varying parallelism or thread count may perturb the output slightly through floating-point reduction order, but the logical computation remains unchanged.

Formally, let $I_Y$ denote the parameters relevant to objective $Y \in \{P,Q\}$. Any parameter relevant to answer quality $Q$ changes the logical computation and must also be considered for serving performance $P$; therefore, \emph{$I_Q \subseteq I_P$}. This yields $\mathcal{X}_{\mathrm{full}} = \mathcal{X}_{\mathrm{algo}} \times \mathcal{X}_{\mathrm{cm}}$, where $\mathcal{X}_{\mathrm{algo}} := \prod_{j \in I_Q} \Theta_j$ contains parameters that affect both $Q$ and $P$, while $\mathcal{X}_{\mathrm{cm}} := \prod_{j \in I_P \setminus I_Q} \Theta_j$ contains parameters that affect only $P$. \ours{} optimizes $\mathcal{X}_{\mathrm{algo}}$ against $(Q,P)$ and delegates $\mathcal{X}_{\mathrm{cm}}$ to \perf{} for $P$ alone.

\begin{table}[h]
  \centering
  \vspace{-1em}
  \caption{System design space searched by \perf{}.}
  \vspace{-1em}
  \Description{A one-column table listing example system design choices.}
  \label{tab:system-design-space}
  \setlength{\tabcolsep}{2pt}
  \begin{tabular}{@{}>{\raggedright\arraybackslash}p{0.20\columnwidth}
                  >{\raggedright\arraybackslash}p{0.36\columnwidth}
                  >{\raggedright\arraybackslash}p{0.40\columnwidth}@{}}
    \toprule
    \textbf{Category} & \textbf{Design choice} & \textbf{Example range} \\
    \midrule
    \multirow{2}{*}{Vector DB} & Search threads & $\{16, 32, 64\}$ \\
     & Parallel mode & \{intra-query, inter-query\} \\
    \midrule
    \multirow{3}{*}{LLM serving} & Parallelism strategy & \{TP, PP, DP, hybrid\} \\
     & GPUs per stage & $\{1, 2, 3, 4\}$ \\
     & Prefill--decode deployment & \{collocated, disaggregated\} \\
    \midrule
    Placement & Stage-to-device mapping & \begin{tabular}[c]{@{}l@{}} Prefill $\to$ GPUs~\{0--7\} \end{tabular} \\
    \midrule
     \multirow{3}{*}{Batching} & Request batch size & $\{1, 2, \ldots, 256\}$ \\
     & Decode batch size & $\{16, 32, 64, 128, 256\}$ \\
     & Dynamic batching wait time & $\{0, 1, 2, 5, 10\}\,\mathrm{ms}$ \\
    \midrule
    \multirow{2}{*}{\begin{tabular}[c]{@{}l@{}}Hardware\\ (Optional)\end{tabular}} & CPU model & \{AMD EPYC, Intel Xeon\} \\
       & GPU inventory & \{4$\times$H100, 8$\times$A100\} \\
    
    \bottomrule
  \end{tabular}
  \vspace{-1em}
\end{table}

\subsubsection{Search-space organization.}
\plan{} parses the user's declarative specification into the two design spaces above. System-only choices in $\mathcal{X}_{\mathrm{cm}}$ are sent to \perf{} (\S\ref{sec:rag-cm}), which exhaustively searches them inside the cost model to return the best predicted performance $\tilde{P}(\mathbf{x})$ for a given algorithm configuration. \plan{} therefore runs the expensive ground-truth optimization only over $\mathcal{X}_{\mathrm{algo}}$, while \perf{} handles the internal system search. It organizes $\mathcal{X}_{\mathrm{algo}}$ as a hierarchical design space (Fig.~\ref{fig:design-space-tree}) governed by two rules: \emph{conditional activation} and \emph{heuristic constraints} that remove invalid or low-value trials. Under conditional activation, a child parameter is active only when its parent choice is active; for example, query-rewriter parameters are searched only if query rewriting is enabled. The heuristic constraints (starred nodes in Fig.~\ref{fig:design-space-tree}) use smooth proxies for derived knobs: IVF sets $\texttt{nlist}=\texttt{factor}\cdot\sqrt{N}$ for corpus size $N$, while PQ selects a valid divisor $M$ of the embedding dimension $D$.

\subsection{The Optimizer}\label{sec:rag-pe:optimizer}
As discussed in \S\ref{sec:bg:limitations}, RAG's algorithm design space is entangled by cross-stage interactions: the best setting for one stage depends on the choices made in the others (Fig.~\ref{fig:bg-interactions}). Optimizing stages independently can therefore miss configurations that work well only as a complete pipeline. \plan{} consequently uses \emph{multi-objective Bayesian optimization} (MOBO) as its global foundation, evaluating and ranking complete configurations against the end-to-end $(Q,P)$ Pareto frontier. MOBO also suits the small evaluation budget: its probabilistic surrogate and acquisition function use prior observations and uncertainty to select the configuration expected to improve the frontier most.

A global optimizer alone, however, sees only the end-to-end objectives and lacks directional information about which stage should change. \plan{} therefore augments the global MOBO with stage-level feedback: stage diagnostics guide exploration toward promising changes, while the global acquisition function still arbitrates complete configurations according to how much they are expected to improve the $(Q,P)$ frontier. Stage information thus directs the search without becoming a separate objective or breaking cross-stage interactions. We first introduce the MOBO foundation and then present our \emph{stage--global co-aware} extensions.

\subsubsection{Preliminary: multi-objective Bayesian optimization.}
\emph{Multi-objective Bayesian optimization} (MOBO) follows the general Pareto-search formulation in \S\ref{sec:bg:pareto-search}. In our setting, $\mathcal{X}=\mathcal{X}_{\mathrm{algo}}$ and $\mathbf{y}(\mathbf{x})=\bigl(P(\mathbf{x}),Q(\mathbf{x})\bigr)$, using the serving-performance and answer-quality objectives defined above. Because evaluating quality is expensive, MOBO seeks to identify the frontier using as few evaluations as possible.

MOBO consists of two components. The first is a cheap probabilistic \emph{surrogate} for each objective---a Gaussian process (GP) fitted to all configurations evaluated so far. Because $\mathcal{X}_{\mathrm{algo}}$ contains many categorical variables and has a hierarchical structure, each GP uses a mixed kernel: a Mat\'ern kernel for the numeric knobs and a Hamming kernel for the categorical ones. Together, they provide an appropriate notion of similarity for this design space~\cite{levesqueBayesianOptimizationConditional2017}.

The second ingredient is an \emph{acquisition function} $\alpha$ that scores any unevaluated configuration by its expected gain in this hypervolume. Let $\mathcal{D}_t$ denote the data after $t$ rounds. The \emph{hypervolume improvement} (HVI) of a candidate outcome $\mathbf{y}$ over $\mathcal{F}$ is $\mathrm{HVI}(\mathbf{y} \mid \mathcal{F}, \mathbf{r}) = \mathrm{HV}(\mathcal{F} \cup \{\mathbf{y}\}; \mathbf{r}) - \mathrm{HV}(\mathcal{F}; \mathbf{r})$. HVI alone cannot score a candidate, because the outcome $\mathbf{y}$ is unknown before the evaluation. \emph{Expected hypervolume improvement} (EHVI) resolves this by averaging HVI over the GP posterior prediction at $\mathbf{x}$, measured against the frontier of the outcomes observed so far~\cite{daultonParallelBayesianOptimization2021}. EHVI, however, trusts those observations to be exact, while our quality scores are noisy: the observed outcomes need not form the true frontier. \emph{Noisy} EHVI (NEHVI) also treats the frontier as uncertain, averaging HVI over posterior draws of the evaluated configurations and the candidate: $\alpha_{\mathrm{NEHVI}}(\mathbf{x}) = \mathbb{E}_{\mathbf{f}\sim p(\mathbf{f}\mid\mathcal{D}_t)}[\mathrm{HVI}(\mathbf{f}(\mathbf{x})\mid\mathcal{F}_{\mathbf{f}},\mathbf{r})]$. We estimate this expectation using $N$ Sobol draws from the joint GP posterior; each draw $\mathbf{f}_j$ induces a frontier $\mathcal{F}_j$ over the evaluated configurations~\cite{daultonParallelBayesianOptimization2021}.

In practice, we maximize \emph{LogNEHVI}, a numerically stabilized variant of NEHVI, using the BoTorch implementation~\cite{amentUnexpectedImprovementsExpected2025,balandatBoTorchFrameworkEfficient2020}. Each round, \plan{} fits the GPs, picks the next configuration by maximizing the acquisition over the whole space, $\mathbf{x}_{t+1} = \operatorname*{arg\,max}_{\mathbf{x}\in\mathcal{X}_{\mathrm{algo}}} \alpha_{\mathrm{LogNEHVI}}(\mathbf{x})$, evaluates it, and appends the result.

These considerations lead us to choose GP+LogNEHVI over SMAC's combination of a random-forest (RF) surrogate and ParEGO~\cite{lindauerSMAC3VersatileBayesian2022,knowlesParEGOHybridAlgorithm2006}. LogNEHVI directly rewards expected improvement in the noisy Pareto hypervolume, whereas ParEGO targets the frontier indirectly through scalarized objectives~\cite{daultonParallelBayesianOptimization2021}. The RF instead learns similarity through tree partitions; its piecewise-constant predictions can create plateaus in the acquisition landscape, leaving little signal for local refinement.

Despite this principled acquisition rule, the standard loop is poorly directed when applied to RAG's large hierarchical design space. The surrogate is fitted only on the two end-to-end objectives, so its posterior uncertainty indicates where the search has not looked, but not \emph{why} a configuration succeeds or which pipeline stage is worth changing. Under a tight quality-evaluation budget, it can therefore spend trials on configurations that are novel yet unlikely to be useful.

\subsubsection{Our optimizer: stage--global co-aware MOBO}
\begin{figure}[t]
  \centering
  \includegraphics[width=\columnwidth]{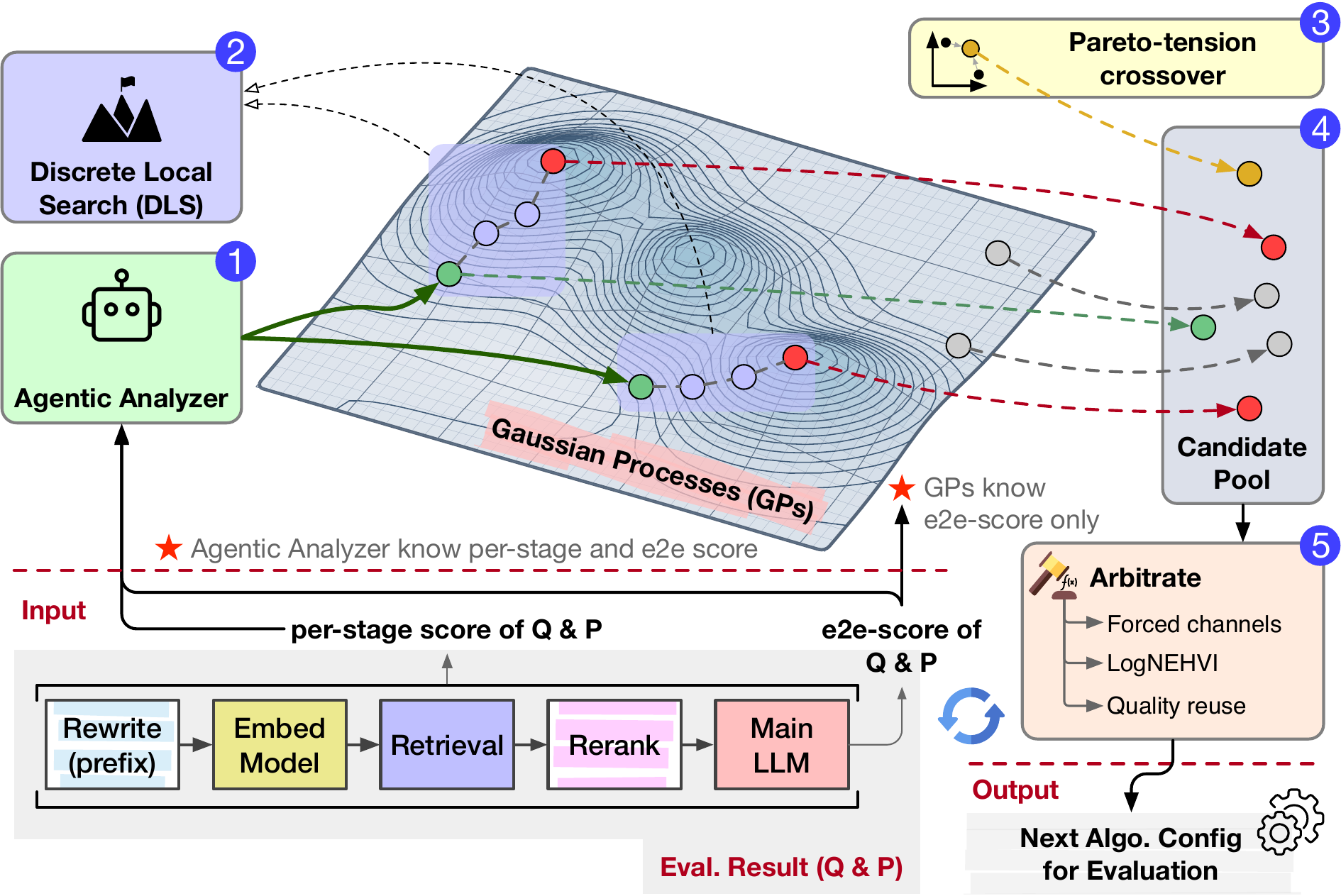}
  \vspace{-2em}
  \caption{The optimizer of \plan{}.}
  \vspace{-2em}
  \label{fig:optimizer}
\end{figure}

\plan{} extends conventional MOBO for RAG with stage-aware diagnostics, heterogeneous candidate generation, and global arbitration, making exploration more directed while preserving end-to-end Pareto optimization. This design is necessary because stage effects are non-separable---the benefit of changing one stage depends on the choices made in the others (Fig.~\ref{fig:bg-interactions}). Greedy stage-wise optimization can therefore miss configurations that work well only as a complete pipeline~\cite{MarkerIncKoreaAutoRAG2026}. Composite-function BO also exploits intermediate outputs, but assumes that the final objective is a known function of those outputs~\cite{astudilloBayesianOptimizationComposite2019b,kudvaMultiObjectiveBayesianOptimization2026}. RAG provides useful per-stage diagnostics, but no known function maps them to end-to-end quality and performance. \plan{} therefore uses these diagnostics to direct candidate generation while leaving complete-configuration comparison to the global MOBO objective.
\plan{} combines three mechanisms, shown in Fig.~\ref{fig:optimizer} and detailed below.

\paragraph{(1) Sub-metric Awareness.} \plan{} optimizes only end-to-end quality $Q$ and performance $P$, while its \emph{agentic analyzer} (\fignum{1}) uses stage-level sub-metrics to diagnose where a configuration falls short. The quality diagnostics are context recall and precision for retrieval and answer faithfulness for the Main LLM; the performance diagnostics are \perf{}'s per-resource capacities $\kappa_r$ (\S\ref{sec:rag-cm:assembly-b}), which expose the bottleneck stage and remaining headroom. We record them as $\mathbf{s}_i$ alongside each observation, yielding $\mathcal{D}_t = \{(\mathbf{x}_i, P_i, Q_i)\}_{i=1}^{t}$ for the GPs and $\mathcal{H}_t = \{(\mathbf{x}_i, P_i, Q_i, \mathbf{s}_i)\}_{i=1}^{t}$ for the analyzer. Reading $\mathcal{H}_t$, the analyzer forms $M_t$ stage-guided candidate configurations $\mathcal{R}^{\mathrm{LLM}}_t=\{\mathbf{c}_1,\ldots,\mathbf{c}_{M_t}\}$. Each $\mathbf{c}_m=\mathbf{b}_m\oplus\boldsymbol{\delta}_m$, $m=1,\ldots,M_t$, is constructed by applying a targeted edit $\boldsymbol{\delta}_m$ to a previously evaluated base $\mathbf{b}_m$ and snapping the result to the valid grid; $\oplus$ overwrites only the edited parameters~\cite{liuLargeLanguageModels2024,laoGPTunerLLMBasedDatabase2025}. Thus, sub-metrics guide candidate generation without becoming surrogate inputs or optimization objectives.

\paragraph{(2) Heterogeneous Candidate Pool.} Standard acquisition optimization uses multiple random restarts, but all candidates are refined against the same surrogate and therefore inherit the same model bias. \plan{} instead constructs $\mathcal{C}_t$ from four complementary channels: Sobol coverage, stage-guided candidates, DLS-based acquisition-guided local refinement, and Pareto-tension crossover (Fig.~\ref{fig:optimizer}).

Two channels determine where to search: \emph{Sobol breadth} draws a hierarchy-aware, space-filling candidate set $\mathcal{B}_t$, while the \emph{agentic analyzer} supplies stage-guided candidates $\mathcal{R}^{\mathrm{LLM}}_t$ (\fignum{1}).

The stage-guided candidates and quasi-random restarts $\mathcal{R}^{\mathrm{rand}}_t$ seed \emph{discrete local search} (DLS, \fignum{2}). Let $\mathbf{z}^{(k)}$ denote the discrete configuration after $k$ DLS steps, and let $\mathcal{N}(\mathbf{z})$ denote its hierarchy-valid one-parameter neighbors. Each $\mathbf{c}_m$ initializes a DLS trajectory at $\mathbf{z}^{(0)}=\mathbf{c}_m$. Writing $\mathcal{N}^{+}(\mathbf{z})=\mathcal{N}(\mathbf{z})\cup\{\mathbf{z}\}$, DLS greedily applies $\mathbf{z}^{(k+1)}=\operatorname*{arg\,max}_{\mathbf{z}\in\mathcal{N}^{+}(\mathbf{z}^{(k)})}\alpha_{\mathrm{LogNEHVI}}(\mathbf{z})$ until convergence.
Thus, the agent chooses a promising region and DLS refines the configuration within it.

\emph{Pareto-tension crossover} (\fignum{3}) starts from the best observed configuration for each objective and applies the single-knob crossover that most strongly moves it toward the other objective.
For $o\in\{P,Q\}$, denote that anchor by $\mathbf{x}_o^\star=\operatorname*{arg\,max}_{\mathbf{x}\in\mathcal{D}_t}o(\mathbf{x})$ and the other objective by $\bar{o}$. Let $\mathbf{x}_{o,j}=\mathbf{x}_o^\star\oplus_j x_{\bar{o},j}^\star$ denote the hierarchy-valid crossover that replaces only knob $j$ with its value from $\mathbf{x}_{\bar{o}}^\star$. It selects $j_o^\star=\operatorname*{arg\,max}_j\tau_{o,j}$, where hats denote min--max normalization and $\tau_{o,j}=\hat{\bar{o}}(\mathbf{x}_{o,j})-\hat{\bar{o}}(\mathbf{x}_o^\star)-[\hat{o}(\mathbf{x}_o^\star)-\hat{o}(\mathbf{x}_{o,j})]_+$, yielding $\mathcal{R}^{\times}_t$. The candidate pool is $\mathcal{C}_t = \mathcal{B}_t \cup \mathcal{R}^{\mathrm{LLM}}_t \cup \mathrm{DLS}\!\left(\mathcal{R}^{\mathrm{LLM}}_t \cup \mathcal{R}^{\mathrm{rand}}_t\right) \cup \mathcal{R}^{\times}_t$, with DLS applied independently to each seed and source labels retained for arbitration (\fignum{4}).

\paragraph{(3) Arbitration.} On ordinary rounds, LogNEHVI scores every candidate in $\mathcal{C}_t$, and \plan{} evaluates the candidate with the largest expected hypervolume improvement, independent of its generator. When the number of consecutive ordinary evaluations with no realized normalized-hypervolume gain reaches a stagnation threshold $\tau_s$, \plan{} activates its \emph{forced channels}. These channels target the largest gap between adjacent points on the normalized realized Pareto frontier and draw candidates from the agent and Pareto-tension crossover. The candidates bypass LogNEHVI and are ranked by the product of their range-normalized posterior standard deviations, with hierarchy-aware novelty breaking ties. Finally, \emph{quality reuse} skips generation and judging when the retrieved passages and quality-relevant downstream configuration match an earlier evaluation: \plan{} reuses its quality score, recomputes performance with \perf{}, records the point, and repeats arbitration.

\section{\ir{}: Intermediate Representation}
\label{sec:rag-ir}

\ir{} bridges quality evaluation and \perf{} through a common workload representation with two properties. It is \emph{system-agnostic}, separating logical work from deployment so \perf{} can search deployment choices and re-cost the same workload on new hardware without rerunning it. It is \emph{RAG-workflow-agnostic}, encoding sequential and agentic pipelines with a common schema so \plan{} can explore workflows while \perf{} evaluates them without workflow-specific modeling logic (Fig.~\ref{fig:overview}).

\ir{} records an order-free workflow schema for throughput and per-request execution traces for latency (\S\ref{sec:rag-cm:assembly-b}).

\textbf{Workflow schema.} An order-free summary of each stage's performance-relevant attributes and aggregate logical work, used by \perf{} to predict throughput.

\textbf{Per-request trace.} For each request $q$, \ir{} records an execution DAG $\mathcal{G}_i^{(q)}$. Each node is one stage invocation and stores its stage type and input/output token counts; edges record dependencies between calls. Repeated calls in iterative or agentic pipelines appear as separate nodes, preserving their order and multiplicity. \ir{} forwards the schema and traces, together with $\mathbf{x}_i$, to \perf{} (\S\ref{sec:rag-cm}).

\section{\perf{}: Cost Model}
\label{sec:rag-cm}

\perf{} is the performance-estimation component of \ours{}. For a fixed RAG algorithm configuration $\mathbf{x}_i$, it estimates the best attainable serving performance on a target system by searching the system design space $\mathcal{X}_{\mathrm{cm}}$, whose parameters affect serving performance but not answer quality. Its inputs are the workload representation produced by \ir{} (a workflow schema and per-request execution traces $\mathcal{G}_i$), the algorithm configuration $\mathbf{x}_i$, and a resource specification of the target system's devices, communication topology, and serving resources.

We denote by $\mathbf{s}\in\mathcal{X}_{\mathrm{cm}}$ one concrete assignment of all system-design parameters (Table~\ref{tab:system-design-space}). For each pair $(\mathbf{x}_i,\mathbf{s})$, \perf{} predicts throughput and end-to-end latency, $\tilde{\mathbf{p}}(\mathbf{x}_i,\mathbf{s})$.

\textbf{Why a cost model for performance.} Predicting performance instead of measuring it serves four roles in \ours{}. \textbf{(R1) A smaller optimizer space.} By moving the system design space into \perf{}, the optimizer only searches the algorithm design space $\mathcal{X}_{\mathrm{algo}}$, so the scarce ground-truth budget is spent only on parameters that affect quality (\S\ref{sec:rag-pe}). \textbf{(R2) RAG serving planning and transfer.} \perf{} lets users plan RAG serving before acquiring the target system and efficiently retarget it to new hardware by re-scoring the serving performance of previously identified Pareto configurations on the new system. \textbf{(R3) Faster optimization iterations.} Because \perf{} predicts performance without deploying each candidate or benchmarking its performance, every optimization iteration avoids this overhead, substantially reducing wall-clock time relative to deployment-based optimizers. \textbf{(R4) Exhaustive system search.} \perf{} exhaustively searches the system design space for each algorithm configuration, yielding higher performance and more effective optimization.

\begin{figure}[t]
  \centering
  \includegraphics[width=\columnwidth]{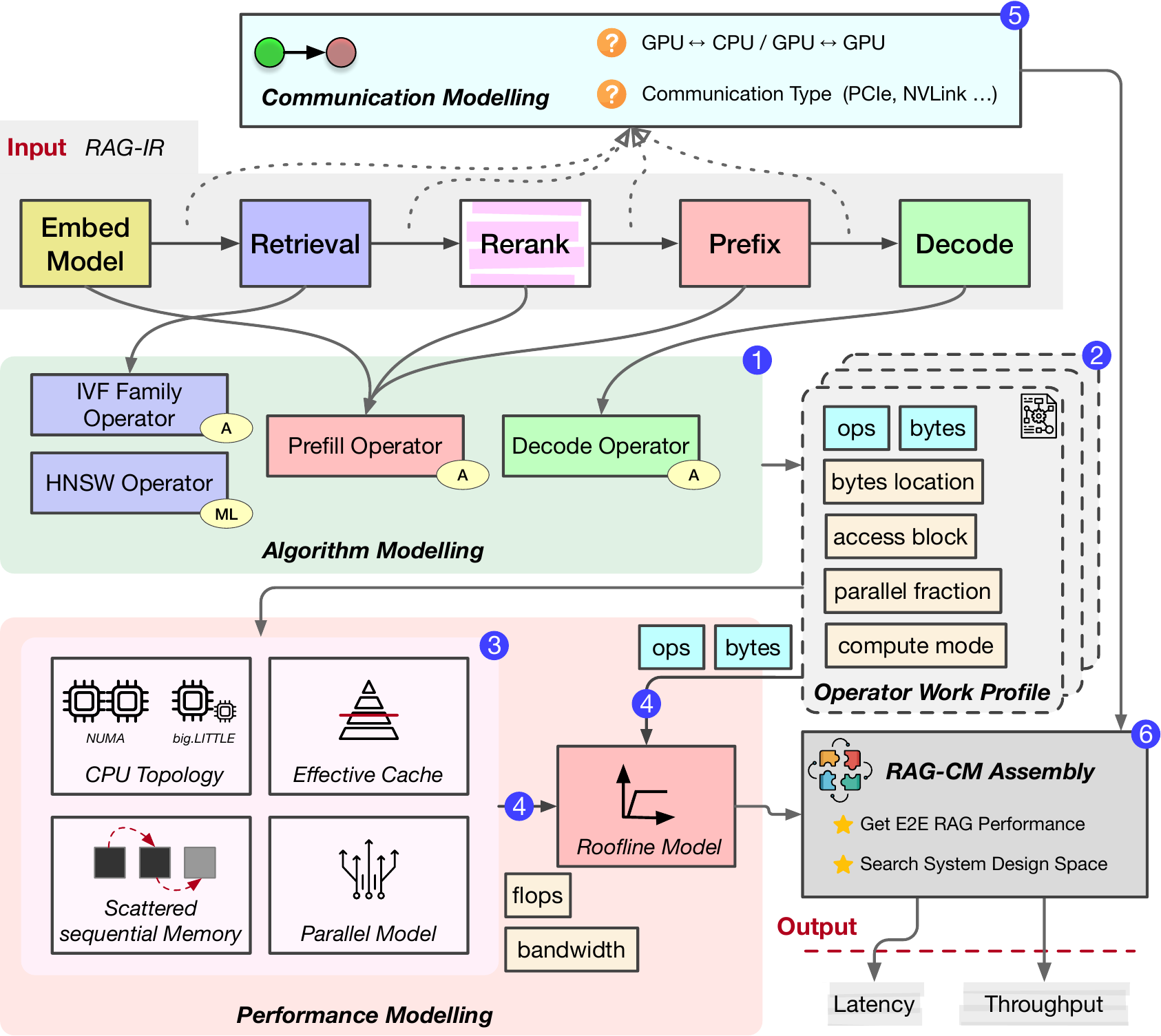}
  \vspace{-1em}
  \caption{Overview of \perf{}. A: analytical; ML: fused machine learning and
  analytical.}
  \vspace{-1em}
  \label{fig:rag-cm-overview}
\end{figure}

\textbf{The four-layer structure.} \perf{} computes $\tilde{\mathbf{p}}$, and hence $\tilde{P}$, in four layers (Fig.~\ref{fig:rag-cm-overview}): an \emph{algorithm} layer that produces one hardware-agnostic Operator Work Profile per RAG stage (\S\ref{sec:rag-cm:algo-b}), a \emph{performance} layer that maps each Operator Work Profile to time on the host (\S\ref{sec:rag-cm:perf-b}), a \emph{communication} layer that prices the data moved between operators (\S\ref{sec:rag-cm:comm-b}), and an \emph{assembly} layer that sweeps the system design space $\mathcal{X}_{\mathrm{cm}}$ and composes the other three layers into the throughput and latency of each algorithm--system configuration pair (\S\ref{sec:rag-cm:assembly-b}).

\subsection{Algorithm Modeling}\label{sec:rag-cm:algo-b}

Given \ir{}'s workflow and execution trace, the algorithm layer models every RAG stage and outputs one hardware-agnostic \emph{Operator Work Profile} per stage---a record of that stage's per-phase operation counts, data movement, and execution characteristics (\fignum{1} in Fig.~\ref{fig:rag-cm-overview}; \S\ref{sec:rag-cm:owp-b}). \perf{} uses GenZ~\cite{bambhaniyaDemystifyingAIPlatform2025} for model-inference operators; the following subsections present its retrieval models, grouped into Clustering Indices (\S\ref{sec:rag-cm:algo-ivf-b}) and Graph Indices (\S\ref{sec:rag-cm:algo-hnsw-b}).

\subsubsection{Clustering Indices}\label{sec:rag-cm:algo-ivf-b}
All three clustering indices share one search skeleton: a query is run through a \emph{coarse quantizer} that selects the $n_{\text{probe}}$ inverted lists nearest to it, the candidate vectors in those lists are \emph{scanned} to compute distances, a \emph{top-$k$ heap} keeps the best candidates, and an optional \emph{re-rank} recomputes exact distances on the survivors. They are \emph{analytical}: given the configuration, the work in every stage is fixed in closed form. Table~\ref{tab:ivf-stages} lists every stage, its compute kernel, the peak it is bounded by, and which variants use it. For most stages, the corresponding Operator Work Profile follows directly from the FAISS source: the ops, bytes, and access pattern of each kernel are fixed by the configuration, so a closed-form formula gives the profile for each such stage listed in Table~\ref{tab:ivf-stages}, with variant-specific formulas where the kernels differ; for example, standard PQ builds its distance table per probed list, whereas Fastscan builds it once per query. Only the scan stage requires special treatment because its work cannot be read directly from the configuration, so we model it below.

\begin{table}[t]
  \centering
  \caption{IVF-family search stages (Flat: IVF-Flat; PQ: IVF-PQ; FS: IVF-PQ Fastscan; FSR: Fastscan with residual).}
  \vspace{-1em}
  \label{tab:ivf-stages}
  \footnotesize
  \setlength{\tabcolsep}{3pt}
  \begin{tabular}{@{}>{\raggedright\arraybackslash}p{0.22\columnwidth}
                    >{\raggedright\arraybackslash}p{0.48\columnwidth}
                    c
                    >{\raggedright\arraybackslash}p{0.155\columnwidth}@{}}
    \toprule
    \textbf{Stage} & \textbf{Compute Mode} & \textbf{Peak} & \textbf{Used by} \\
    \midrule
    1. Coarse quantizer   & BLAS GEMM if per-thread batch $\ge 20$, else hand-vectorized SIMD distance kernel & FP32 & all \\
    \midrule
    2. LUT build          & non-residual, built once/query (FS); residual-dependent (PQ, FSR) & FP32 & PQ, FS, FSR \\
    \midrule
    3a. PQ scan            & ADC table lookup (PQ) / \texttt{vpshufb} (FS, FSR)     & int  & PQ, FS, FSR \\
    \midrule
    3b. Flat scan          & full-vector SIMD L2                                    & int  & Flat \\
    \midrule
    4. Top-$k$ heap       & branch-heavy integer                                   & int  & all \\
    \midrule
    5. Refinement         & exact SIMD L2 on candidates                            & FP32 & optional \\
    \bottomrule
  \end{tabular}
  \vspace{-2em}
\end{table}

\paragraph{PQ/flat scan: data-aware imbalance.} The scan's cost is driven by $N_{\text{scan}}$, the number of database vectors a query touches---a vector \emph{count}, not an op or byte total. Under a fixed PQ configuration, scoring each scanned vector requires reading its $b_{\text{code}}$-byte PQ code and summing one precomputed distance-table entry per subquantizer, so the scan's ops are $N_{\text{scan}}$ times the per-vector work and its bytes are $N_{\text{scan}}\,b_{\text{code}}$; this is how $N_{\text{scan}}$ sets the scan operator's profile. IVF makes $N_{\text{scan}}$ far smaller than the corpus size $N$: a query scans only the $n_{\text{probe}}$ of $n_{\text{list}}$ inverted lists nearest to it, a fraction $\rho = n_{\text{probe}}/n_{\text{list}}$ of all vectors. The naive estimate $N_{\text{scan}}=\rho N$ assumes equal-sized lists, but real embeddings cluster, so a query's nearest lists are larger than average and it scans more; we correct with a cell-imbalance factor $f$ measured from the built index, so that $N_{\text{scan}} = f(n_{\text{probe}})\,\rho\,N$, where $f$ is the ratio of the vectors a real query actually scans---obtained from the index's per-list sizes and the nearest-list assignments of a query sample---to the balanced $\rho N$. This $f$ is the only index-dependent input the model needs, and it is a pure property of the corpus and embedding---never of the hardware---while the remaining index knobs enter the cost model as analytical parameters. \perf{} therefore caches $f$ once per (corpus, embedding)---only a handful of combinations across a whole design space---and from then on predicts the scan count for any configuration without ever building an index again.

\subsubsection{Graph Indices}\label{sec:rag-cm:algo-hnsw-b}
HNSW is \emph{ML-analytical}. A query first descends greedily through the $L\approx\log_M N$ upper layers---a few hops each over $M$-neighbor lists---to reach a good entry point, then runs a bounded best-first (beam) search of width $\mathit{ef}_s$ at the base layer, which performs almost all of the work. Where IVF's scan count follows in closed form from the configuration, the size of the base-layer neighborhood an HNSW query explores is data-dependent---it grows with the data's intrinsic geometry and shrinks on clustered corpora where the beam converges fast---and has no closed form. The model therefore \emph{predicts} the two quantities that drive the Operator Work Profile: the per-query distance computations $n_{\text{dis}}$ and graph hops $n_{\text{hops}}$. These fold into the profile's $W$ and $V$ exactly as IVF's $N_{\text{scan}}$ does (\S\ref{sec:rag-cm:owp-b})---$n_{\text{dis}}$ setting the dominant base-layer FLOPs and random-vector reads, and $n_{\text{hops}}$ the visited-set and heap overhead. Concretely, the HNSW operator model estimates $\hat{n}_{\text{dis}} = c_M\,\hat{g}_{\text{dis}}(\mathbf{z})$ and $\hat{n}_{\text{hops}} = \hat{g}_{\text{hops}}(\mathbf{z})$ with feature vector $\mathbf{z} = (d, M, \mathit{ef}_s, \mathrm{LID})$: a learned predictor $\hat{g}$ emits the workload counts $\hat{\mathbf{n}}=(\hat{n}_{\text{dis}},\hat{n}_{\text{hops}})$, and a factor $c_M$ obtained through one-time per-corpus calibration fixes the scale of $\hat{n}_{\text{dis}}$. These counts determine the HNSW Operator Work Profile; the following paragraph defines $\hat{g}$ and its lightweight calibration.

\paragraph{Workload predictor.} The predictor $\hat{g}$ is two independent gradient-boosted regressors, $\hat{g}_{\text{dis}}$ and $\hat{g}_{\text{hops}}$; both are predicted because their ratio is not constant---it drifts with $\mathit{ef}_s$---and each contributes a different part of the Operator Work Profile. The lone data-dependent feature is the corpus local intrinsic dimensionality $\mathrm{LID}$, estimated by an Amsaleg MLE on a query sample; it carries the geometry that lets a single model generalize across corpora---from high-LID synthetic training data to the low-LID, tightly clustered distributions of real image and text embeddings, on which a configuration-only estimate over-counts because the beam terminates sooner than the ambient $d$ suggests. Crucially, every feature in $\mathbf{z}$ is either a search-space knob or a property of the corpus itself: the predictor uses \emph{no} statistic of a constructed graph (degree, layer counts), so it scores candidate configurations whose indexes have not been built---exactly the regime the optimizer explores. The predictor requires only a one-time calibration for each corpus, which is reused across all unbuilt configurations and therefore adds little overhead.

\subsection{Operator Work Profile}\label{sec:rag-cm:owp-b}
Each operator emits a hardware-agnostic \emph{Operator Work Profile}, the interface between the algorithm and performance layers. Computed once per operator phase, it contains \textbf{work} $(W,V)$---the operation count and bytes moved after cache reuse---and four \textbf{execution characteristics}: (1)~the \emph{compute mode} (e.g.\ \textsc{blas}/\textsc{simd}), which selects compute efficiency; (2)~the \emph{bytes location}, determined by the working-set size and unique byte volume $V_{\text{uniq}}\!\le\!V$, which selects the memory level; (3)~the \emph{access-block size}, which distinguishes sequential from scattered bandwidth; and (4)~the \emph{parallel fraction} for Amdahl scaling. The performance layer maps these fields to effective compute and memory rates $(\pi_{\text{eff}},\beta_{\text{eff}})$. Data-dependent corrections modify $W$ and $V$ rather than the profile schema, allowing one performance model to serve all operators.

\subsection{Performance Modeling}\label{sec:rag-cm:perf-b}
The performance layer maps an \emph{Operator Work Profile} to wall-clock time in two composable steps (\fignum{3},\,\fignum{4} in Fig.~\ref{fig:rag-cm-overview}): a \emph{roofline} bounds each phase by the slower of its compute and memory time, and \emph{Amdahl's law} composes the per-phase times across threads under the chosen parallel mode. The intermediate points below exist only to feed these two: the hardware model supplies the roofline's two rates ($\pi_{\text{eff}}$, $\beta_{\text{eff}}$), and the parallel model sets how the per-phase results combine into end-to-end latency and throughput.

\paragraph{Roofline kernel.} For a phase with $W$ operations and data volume $V$, the model predicts $T = T_{\mathrm{ovh}} + \max\!\left(\frac{W}{\pi_{\mathrm{eff}}},\frac{V}{\beta_{\mathrm{eff}}}\right)$, where $\pi_{\mathrm{eff}}$ and $\beta_{\mathrm{eff}}$ are the effective compute throughput and bandwidth, and $T_{\mathrm{ovh}}$ captures fixed per-call costs.

\paragraph{Hardware and execution factors.} To derive these effective rates and compose phase times, \perf{} accounts for (1)~\emph{CPU topology}, including heterogeneous cores and NUMA effects; (2)~\emph{effective cache capacity} under sharing among concurrent threads; (3)~\emph{scattered--sequential memory access}, which is sequential within a block but random across blocks; and (4)~\emph{parallel execution}, using separate compute and memory thread counts with Amdahl scaling.

\subsection{Communication Modeling}\label{sec:rag-cm:comm-b}
Given a deployment, \perf{}'s communication model estimates the time to move every inter-stage and intra-LLM payload across its selected devices (\fignum{5} in Fig.~\ref{fig:rag-cm-overview}). It derives these costs from the concrete topology: every GPU pair carries its own bandwidth and startup latency $(\beta_{ij},\ell_{ij})$, and GPU--CPU hops use a host-transfer class---covering tensor-parallel (TP) and pipeline-parallel (PP) collectives, data-parallel (DP) replicas, GPU--GPU stage handoffs such as the prefill$\to$decode KV-cache transfer under 1P/1D disaggregation, and CPU-side boundaries. A cross-device edge $e$ sums its fastest endpoint-disjoint pair bandwidths into $\beta_e$ and takes its slowest startup as $\ell_e$; LLM-internal TP/PP traffic folds into the prefill/decode service time rather than adding assembly edges; and each explicit cross-device edge is priced as $\ell_e+M_e/\beta_e$ over its payload $M_e$, with text and dataframe boundaries (e.g., retrieval$\to$reranker) priced as host transfers.

\subsection{\perf{} Assembly}
\label{sec:rag-cm:assembly-b}
The assembly layer (\fignum{6} in Fig.~\ref{fig:rag-cm-overview}) has two functions: it enumerates and materializes each $\mathbf{s}\in\mathcal{X}_{\mathrm{cm}}$ as one deployment and execution plan, then composes stage-level operator and communication predictions into steady-state, end-to-end serving performance. For each candidate it prices, assembly predicts the saturation throughput $\tilde{T}(\mathbf{s})$ and the corresponding mean latency $\tilde{L}(\mathbf{s})$ once serving reaches steady state; startup behavior is outside the model. \perf{} uses the performance objective and SLOs in the user's optimization target to select among these configurations and exposes the selected performance to \plan{} as $\tilde{P}(\mathbf{x}_i)$.

\subsubsection{Steady-state performance under saturated closed-loop serving} We model a saturated closed-loop deployment with a fixed number of concurrent clients, each issuing its next query after the previous response. The resulting $\tilde{T}$ is the maximum sustainable throughput, and $\tilde{L}$ is the steady-state response latency at that operating point. This target characterizes the deployment's attainable capacity without introducing an external arrival rate. An open-system latency instead depends on the offered load and can make the same deployment appear lightly loaded, near saturation, or unstable.

\subsubsection{Trace-driven assembly} For each algorithm--system pair $(\mathbf{x}_i,\mathbf{s})$, \perf{} instantiates a closed discrete-event model from \ir{}'s workload representation and the operator and communication models (\S\ref{sec:rag-cm:comm-b}). The simulation captures four mechanisms: (1)~\textbf{Continuous and dynamic batching:} LLM stages use continuous batching, while other stages dispatch batches by size or timeout; (2)~\textbf{Automatic prefix caching:} repeated calls prefill only the uncached prompt suffix within KV-cache capacity; (3)~\textbf{Decoupled batch limits:} generator decode and other stages use independent batch-size limits~\cite{jiangRAGOSystematicPerformance2025}; and (4)~\textbf{Shared-resource contention:} collocated stages time-share GPUs and CPUs. At concurrency $N$, $\widehat{T}(\mathbf{s},N)$ and $\widehat{L}(\mathbf{s},N)$ denote steady-state throughput (QPS) and mean end-to-end latency, respectively. To scale to thousands of candidates, \perf{} uses a cheap analytical model to shortlist the best batch settings per deployment topology, then simulates only those candidates to throughput saturation and reports only simulation results.

\section{Evaluation}
\label{sec:eval}

Our evaluation is organized around four questions.

\textbf{RQ1 (End-to-end value).} Across random seeds and under the same evaluation budget, does \ours{} consistently find better quality--performance Pareto frontiers than existing optimization methods?

\textbf{RQ2 (Optimizer ablation).} With the rest of \ours{} held fixed, does the optimizer inside \plan{} outperform existing alternatives?

\textbf{RQ3 (Cost-model accuracy).} How accurate is \perf{}, in absolute error and, more importantly, in ranking candidate deployments?

\textbf{RQ4 (System transfer).} After migrating to new hardware, how much does \ours{} improve normalized hypervolume over re-optimizing from scratch under the same evaluation budget?

\subsection{Experimental Setup}
\label{sec:eval:setup}

\textbf{Hardware.} We use two NUMA servers as measured deployment targets with different CPU and GPU resources: \textbf{SysA} has two AMD EPYC 9124 sockets, 32 CPU cores in total (16 cores per socket, 3.0~GHz base and up to 3.7~GHz boost), two NUMA nodes, and four NVIDIA H100 GPUs; \textbf{SysB} has two AMD EPYC 7742 sockets, 128 CPU cores in total (64 cores per socket, 2.25~GHz base and up to 3.4~GHz boost), two NUMA nodes, and eight NVIDIA A100 80GB GPUs.

\textbf{Datasets and Quality Metrics.} For the end-to-end evaluation, we use 100 queries each from RAGEval~\cite{zhuRAGEvalScenarioSpecific2025} to avoid contamination from LLM training data~\cite{xuWizardLMEmpoweringLarge2025,qiLong$2$RAGEvaluatingLongContext2025}, and MS MARCO~\cite{bajajMSMARCOHuman2018} to evaluate on a larger, more diverse dataset. The \perf{} accuracy evaluation uses additional datasets, which we describe in \S\ref{sec:eval:cm-accuracy}. The quality objective $Q$ is RAGAS answer correctness~\cite{esRAGAsAutomatedEvaluation2024}, which scores factual agreement with the gold answer rather than lexical overlap---so valid paraphrases are not penalized---in $[0,1]$, averaged over queries. As per-stage diagnostics, \plan{} reads three RAGAS sub-metrics (\S\ref{sec:rag-pe:optimizer}): context recall, context precision, and faithfulness. All LLM-as-judge scoring uses DeepSeek-V4-Flash.

\textbf{Algorithm Design Spaces.} Our algorithm design space follows Fig.~\ref{fig:design-space-tree}; here, we specify only the choices not detailed in the figure, while all other components and parameter ranges remain as shown. For RAGEval, the chunk size is chosen from $\{128,\allowbreak 256,\allowbreak 512,\allowbreak 1024,\allowbreak 2048\}$; the embedding model is BGE-small~\cite{FlagOpenFlagEmbedding2026} (384-d), all-mpnet-base-v2~\cite{songMPNetMaskedPermuted2020} (768-d), or BGE-M3~\cite{chenM3EmbeddingMultiLingualityMultiFunctionality2024} (1024-d); the reranker is TART~\cite{asaiTaskawareRetrievalInstructions2023}, ColBERT~\cite{khattabColBERTEfficientEffective2020a}, SentenceTransformer~\cite{reimersSentenceBERTSentenceEmbeddings2019}, or FlagEmbedding~\cite{FlagOpenFlagEmbedding2026, chenM3EmbeddingMultiLingualityMultiFunctionality2024}; and the query rewriter and main LLM independently use a Qwen2.5~\cite{baiQwenTechnicalReport2023} model from $\{1.5,\allowbreak 3,\allowbreak 7,\allowbreak 14\}$B. For MS MARCO, we fix the chunk size at 2048 and the embedding model to all-mpnet-base-v2 to keep indexing its 8.84M passages tractable.

\textbf{System Design Spaces.} We instantiate the system axes of Table~\ref{tab:system-design-space}. The end-to-end and optimizer-ablation experiments (\S\ref{sec:eval:e2e}, \S\ref{sec:eval:optimizer-ablation}) share the four-H100 SysA space: \perf{} maximizes throughput over collocated and disaggregated GPU-stage placements, all feasible stage-to-GPU mappings and LLM parallelism plans (at most four GPUs per stage), request and decode batch sizes in $\{1,2,4,\ldots,256\}$ with $B_{\mathrm{decode}}\ge B_{\mathrm{request}}$, and dynamic-batching waits in $\{2,10,50\}\,\mathrm{ms}$; FAISS threads follow $\min(\text{retrieval batch},\text{physical cores})$ under inter-query parallelism. For transfer to the eight-A100 SysB (\S7.5), only the GPU budget and per-stage cap grow from four to eight.

\emph{Baseline system space.} The baselines in \S\ref{sec:eval:e2e} have no performance model, so a fair comparison lets them search the algorithm design space jointly with the system axes as one entangled space. But every candidate they propose must be deployed and measured, and the full system design space would spend most of this budget on poor system variants. We therefore compress all system axes that \perf{} searches into two categorical choices, expert-curated to be strong on SysA: eight \emph{deployment presets} (collocated TP/PP plans and 1P/1D--2P/2D prefill/decode disaggregation, with non-main-LLM stages either sharing GPUs with the main LLM or using GPUs left unused by it) crossed with six $B_{\mathrm{request}}{\times}B_{\mathrm{decode}}$ \emph{batch presets}.

\subsection{End-to-End Optimization}
\label{sec:eval:e2e}

\begin{figure}[t]
  \centering
  \includegraphics[width=\columnwidth]{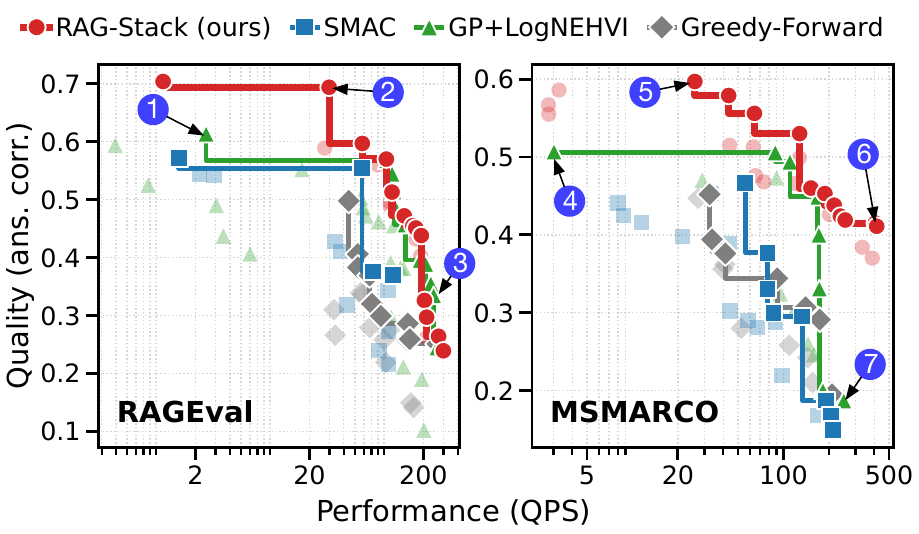}
  \vspace{-2em}
  \caption{End-to-end quality--performance Pareto frontiers (log-scaled performance). Translucent markers show per-seed frontiers; each line pools seeds 43--45. Selected \ours{} configurations are re-measured on SysA.}
  \vspace{-1em}
  \label{fig:eval-e2e-pareto}
\end{figure}

Figure~\ref{fig:eval-e2e-pareto} compares the measured quality--throughput Pareto frontiers under the protocol in \S\ref{sec:eval:setup}. Following \S\ref{sec:bg:pareto-search}, we compute hypervolume from linearly normalized quality and raw throughput per dataset; throughput is log-scaled only in the plots for readability. We then compute normalized hypervolume independently for each seed and report the three-seed mean. Against GP+LogNEHVI~\cite{daultonParallelBayesianOptimization2021}, the strongest baseline on both datasets, \ours{} averages 0.514 versus 0.364 on RAGEval and 0.635 versus 0.259 on MS MARCO; the corresponding per-seed relative improvements average 52.5\% and 153.2\%, respectively. \ours{}'s per-seed hypervolume exceeds the strongest baseline's in all six seed--dataset pairs.

As a secondary benefit, \ours{} completes the end-to-end search in 4.59 hours on RAGEval and 3.60 hours on MS MARCO, faster than all baselines except Greedy-Forward (Table~\ref{tab:eval-baseline-execution-time}). This efficiency holds even though \perf{} evaluates 570--5,400 system configurations per candidate, because it prices them with the model rather than deploying each one.

\begin{table}[t]
  \centering
  \caption{Mean end-to-end cold-start runtime on SysA.}
  \vspace{-1em}
  \label{tab:eval-baseline-execution-time}
  \small
  \setlength{\tabcolsep}{4pt}
  \begin{tabular*}{\columnwidth}{@{\extracolsep{\fill}}lrr@{}}
    \toprule
    Optimizer & RAGEval (hours) & MS MARCO (hours) \\
    \midrule
    Greedy-Forward~\cite{MarkerIncKoreaAutoRAG2026}
      & $2.25 \pm 0.23$ & $3.16 \pm 0.06$ \\
    GP+LogNEHVI~\cite{daultonParallelBayesianOptimization2021}
      & $5.60 \pm 0.40$ & $5.68 \pm 0.51$ \\
    SMAC~\cite{lindauerSMAC3VersatileBayesian2022}
      & $8.03 \pm 0.42$ & $7.03 \pm 0.37$ \\
    \midrule
    \ours{} (ours)
      & $4.59 \pm 0.51$ & $3.60 \pm 1.79$ \\
    \bottomrule
  \end{tabular*}
  \vspace{-2em}
\end{table}

To explain these trade-offs, we inspect representative RAG configurations in Figure~\ref{fig:eval-e2e-pareto} and attribute their gains to algorithm and system choices.

On RAGEval (left), \ours{} point~\fignum{2} dominates the baseline-best point~\fignum{1} with higher quality at $81\times$ the throughput. Both use a 14B generator, but \fignum{1} adds 3B multi-query expansion to HNSW ($\mathit{ef}_s{=}128$, top-$k{=}16$) without reranking and uses a fixed 1P/2D plan (one prefill and two decode replicas, batch $16{\times}32$, with 14B prefill on one H100). \fignum{2} disables expansion and uses 2048-character chunks with ColBERT after top-$k{=}8$ IVF-PQ to rank the retrieved documents by relevance, raising answer correctness from 0.594 to 0.638, while \perf{} selects a faster serving plan.

GP+LogNEHVI is seed-dependent: only a few of its points rise above \ours{}'s frontier, all of them from seed 43 and confined to narrow throughput ranges, the widest at \fignum{3} (245.6 vs.\ 204.1~QPS; 0.335 vs.\ 0.326 quality). \ours{} still has higher seed-43 hypervolume, and SMAC and Greedy-Forward never win on both axes. \fignum{3} replaces \ours{}'s top-$k{=}1$ plus ColBERT with reranker-free top-$k{=}4$ retrieval over shorter chunks, then adds 1P/2D disaggregation and a larger batch. Once seed 43 finds this region, LogNEHVI concentrates later evaluations nearby; the other seeds never enter it---stochastic discovery followed by exploitation, not a stable baseline advantage.

On MS MARCO (right), \ours{} point~\fignum{5} dominates the quality-best baseline point~\fignum{4} with globally best quality at $8.6\times$ the throughput: \fignum{4} relies on top-$k{=}64$, whereas \fignum{5} uses top-$k{=}16$ and a \perf{}-selected disaggregated layout. At the throughput extreme, \fignum{6} dominates the baselines' fastest point~\fignum{7} with $1.65\times$ the throughput and $2.2\times$ the quality. Both use a 1.5B generator; \fignum{6} retains accurate, uncompressed HNSW and cuts only to top-$k{=}1$---sufficient because each fixed 2048-character passage forms one chunk---then parallelizes generation with TP2$\times$PP2 and encoding with TP4. \fignum{7} instead combines IVF-PQ-FS with ReAct, whose retrieval errors the small generator cannot correct.

Across the overlapping quality range in Figure~\ref{fig:eval-e2e-pareto}, serving at equal quality is $1.5$--$7.5\times$ cheaper on MS MARCO than on RAGEval, likely due to public-benchmark leakage into LLM training data~\cite{xuWizardLMEmpoweringLarge2025,qiLong$2$RAGEvaluatingLongContext2025}. Public benchmarks may therefore understate the serving cost of quality, motivating RAGEval as a fresh-data control.

\subsection{Optimizer Ablation}
\label{sec:eval:optimizer-ablation}

\begin{figure}[t]
  \centering
  \includegraphics[width=\columnwidth]{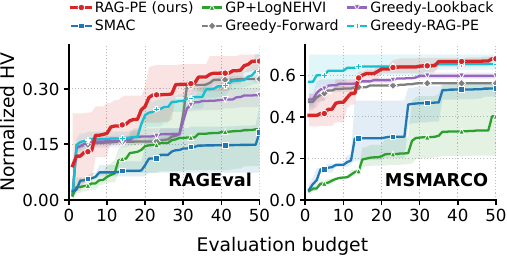}
  \vspace{-2em}
  \caption{Optimizer ablation: dominated hypervolume versus evaluation budget, holding the rest of \ours{} fixed and varying only the optimizer inside \plan{}. Lines are means over 3 seeds; bands show seed min/max.}
  \vspace{-2em}
  \label{fig:eval-optimizer-hv}
\end{figure}

To isolate the contribution of \plan{}'s optimizer, we keep every other part of \ours{} unchanged and replace only this optimizer with each baseline in turn. We compare \plan{} with SMAC~\cite{lindauerSMAC3VersatileBayesian2022}, GP+LogNEHVI~\cite{balandatBoTorchFrameworkEfficient2020}, Greedy-Forward, and Greedy-Lookback~\cite{MarkerIncKoreaAutoRAG2026}; the last alternates backward and forward passes after its initial forward pass. Greedy-\plan{} is a hybrid that switches from Greedy-Forward to \plan{} after 30 of 50 evaluations. Each method runs with seeds 43--45, and Figure~\ref{fig:eval-optimizer-hv} reports mean normalized hypervolume by budget.

\plan{} is the most consistent optimizer across both datasets: against the four standalone baselines, it attains the highest mean hypervolume at 77 of 80 post-warm-start budgets. At budget 50, it reaches 0.375 versus Greedy-Forward's 0.327 on RAGEval and 0.681 versus Greedy-Lookback's 0.599 on MS MARCO, relative gains of 14.8\% and 13.8\%. No single baseline is consistently strongest: the strongest method differs between the two datasets in this ablation, while the end-to-end comparison identifies a different strongest baseline (\S\ref{sec:eval:e2e}). In contrast, \plan{} remains strong across settings and contributes independently to \ours{}'s end-to-end advantage..

\begin{figure}[t]
  \centering
  \includegraphics[width=\columnwidth]{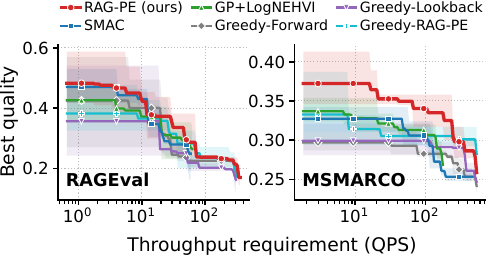}
  \vspace{-2em}
  \caption{Best attainable quality versus throughput requirement: per optimizer, the highest-quality configuration meeting the requirement within the budget (lines: mean over 3 seeds; bands: seed min/max).}
  \vspace{-1em}
  \label{fig:eval-constrained-quality}
\end{figure}

Because dominated hypervolume also credits configurations in regions no deployment would use (extremely low quality or throughput), Figure~\ref{fig:eval-constrained-quality} re-reads the same archives from a deployment perspective: the best quality each optimizer can serve once a minimum throughput is required.

\begin{figure}[t]
  \centering
  \includegraphics[width=\columnwidth]{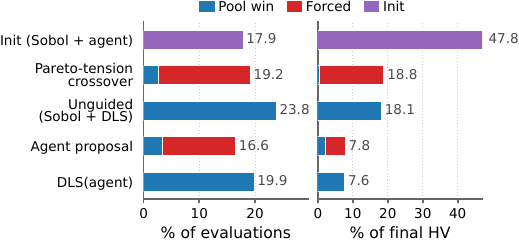}
  \vspace{-2em}
  \caption{Per-channel contributions to \plan{}'s results in \S\ref{sec:eval:optimizer-ablation}: shares of all quality evaluations (left) and final hypervolume (right).}
  \vspace{-1em}
  \label{fig:eval-channel-efficiency}
\end{figure}

Figure~\ref{fig:eval-channel-efficiency} shows that initialization is critical: the Sobol-plus-agent warm start accounts for the largest share of the final hypervolume. The remaining channels also make non-trivial contributions to the final hypervolume.

\subsection{\perf{} Accuracy}
\label{sec:eval:cm-accuracy}

\begin{table}[t]
  \centering
  \caption{\perf{} prediction accuracy. Error is MAPE; $\rho_s$ and $r$ denote Spearman and Pearson correlation; $n$ is the number of configurations.}
  \vspace{-1em}
  \label{tab:eval-rag-cm-accuracy}
  \small
  \setlength{\tabcolsep}{2pt}
  \begin{tabular*}{\columnwidth}{@{\extracolsep{\fill}}lrrrrrrr@{}}
    \toprule
    & & \multicolumn{3}{c}{SysA} & \multicolumn{3}{c}{SysB} \\
    \cmidrule(lr){3-5}\cmidrule(l){6-8}
    Component & $n$ & Err. $\downarrow$ & $\rho_s \uparrow$ & $r \uparrow$
              & Err. $\downarrow$ & $\rho_s \uparrow$ & $r \uparrow$ \\
    \midrule
    IVF-Flat (Latency)  & 137  & 6.7\%  & 0.998 & 1.000 & 11.4\% & 0.993 & 0.997 \\
    IVF-PQ (Latency)    & 137  & 11.3\% & 0.996 & 0.997 & 13.2\% & 0.996 & 0.998 \\
    IVF-PQ-FS (Latency) & 137  & 17.0\% & 0.989 & 0.997 & 12.8\% & 0.994 & 0.998 \\
    HNSW (Latency)      & 1080 & 17.0\% & 0.986 & 0.971 & 20.1\% & 0.977 & 0.957 \\
    \midrule
    Sequential RAG (Latency) & 201  & 11.9\% & 0.973 & 0.955 & 13.4\% & 0.965 & 0.950 \\
    Sequential RAG (QPS)     & 201  & 11.0\% & 0.975 & 0.969 & 14.6\% & 0.965 & 0.950 \\
    Agentic RAG (Latency)    & 30   & 8.1\%  & 0.966 & 0.987 & 13.8\% & 0.984 & 0.979 \\
    Agentic RAG (QPS)        & 30   & 12.2\% & 0.969 & 0.974 & 20.3\% & 0.967 & 0.980 \\
    \addlinespace[1pt]
    \textbf{RAG Overall (Latency)} & \textbf{231} & \textbf{11.4\%} & \textbf{0.978} & \textbf{0.958} & \textbf{13.5\%} & \textbf{0.970} & \textbf{0.952} \\
    \textbf{RAG Overall (QPS)}     & \textbf{231} & \textbf{11.2\%} & \textbf{0.978} & \textbf{0.952} & \textbf{15.3\%} & \textbf{0.968} & \textbf{0.952} \\
    \bottomrule
  \end{tabular*}
\end{table}

Table~\ref{tab:eval-rag-cm-accuracy} evaluates calibrated \perf{} predictions on SysA and SysB at the operator and RAG pipeline levels. The operator study spans four vector-search corpora (SIFT1M~\cite{jegouProductQuantizationNearest2011}, GloVe, GIST1M~\cite{jegouProductQuantizationNearest2011}, and Deep1M~\cite{yandexEfficientIndexingBillionScale2016}) and two embedded RAG corpora (ELI5~\cite{fanELI5LongForm2019} and TriviaQA~\cite{joshiTriviaQALargeScale2017}), varying batch size and CPU parallelism across HNSW and IVF variants (full grids in the artifact). The RAG pipeline study crosses 77 serving configurations with RAGEval~\cite{zhuRAGEvalScenarioSpecific2025}, ELI5, and TriviaQA, covering sequential and ReAct-style agentic workflows and varying models, indexes, batching, parallelism, colocation, disaggregation, and dynamic batching; each deployment is compared against measured saturated closed-loop QPS and mean end-to-end latency.

For guiding system space search, preserving the relative order of configurations matters more than minimizing absolute prediction error. Across both systems, calibrated \perf{} attains Spearman correlations of 0.968--0.978 for latency and QPS, while keeping overall MAPE at 11.2--15.3\%. Even for random-access-heavy IVF-PQ-FS and HNSW, where MAPE rises to 20.1\%, Spearman correlation remains at least 0.977. Thus, the residual prediction error largely preserves configuration ordering, providing the ranking signal that \plan{} needs.

\subsection{System Transfer}
\label{sec:eval:transfer}

\begin{figure}[t]
  \centering
  \includegraphics[width=\columnwidth]{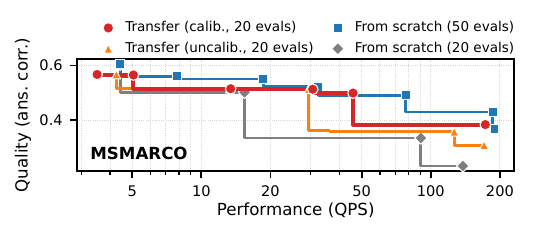}
  \vspace{-2em}
  \caption{System transfer from SysA to SysB. The transferred Pareto configurations are re-measured end-to-end on SysB; all points shown are measured.}
  \vspace{-2em}
  \label{fig:eval-system-transfer}
\end{figure}

We evaluate system transfer by migrating optimized RAG configurations from SysA to SysB. \ours{} re-scores the existing quality archive with \perf{} instantiated for SysB, spends 20 additional evaluations refining the frontier, and re-measures the selected configurations on SysB. The \emph{uncalibrated} variant uses only SysB's hardware description, whereas the \emph{calibrated} variant additionally calibrates \perf{} on the target machine. Figure~\ref{fig:eval-system-transfer} compares both variants with from-scratch optimization on SysB under 20- and 50-evaluation budgets.

With the same 20-evaluation budget, uncalibrated and calibrated transfer improve normalized HV over from-scratch optimization by 108.4\% (0.331 vs. 0.159) and 182.2\% (0.448 vs. 0.159), respectively.

\section{Limitations and Discussion}
\label{sec:limitations}

Our current support for agentic RAG is trace-driven: \ir{} must execute the pipeline to observe its runtime control flow before \perf{} can predict its performance. Consequently, \plan{} cannot densely sample the performance objective before quality evaluation---which would let a multi-task Gaussian process (MTGP) exploit abundant performance-only observations to improve sample efficiency---or filter SLO-violating configurations before they enter the candidate pool. We have implemented trace-free static \ir{} and \perf{} paths for sequential RAG, whose control flow is known from the configuration alone; extending such pre-execution performance prediction to agentic RAG remains future work.

\section{Conclusion}
\label{sec:conclusion}

We present \ours{}, a system for efficiently discovering quality--performance Pareto frontiers across RAG algorithms and serving systems. \ours{} combines \plan{} for sub-metric-aware multi-objective exploration, \ir{} for representing executed sequential and agentic workflows, and \perf{} for predicting serving performance and searching system configurations without deploying every candidate. Experiments show that \ours{} improves normalized hypervolume over the strongest end-to-end baseline, with three-seed mean gains of 52.5\% on RAGEval and 153.2\% on MS MARCO.

\begin{acks}
We sincerely thank Yongjun He and Gustavo Alonso for their insightful discussions and invaluable suggestions throughout the development of this work.
\end{acks}

\bibliographystyle{ACM-Reference-Format}
\bibliography{references}

\end{document}